\documentclass[twocolumn,tighten]{aastex62}
\usepackage{amsmath,amssymb,amstext}
\usepackage{apjfonts}
\allowdisplaybreaks

\newcommand{\avg}[1]{\left\langle{#1}\right\rangle}
\DeclareMathOperator\erf{erf}
\usepackage{tikz}
\usetikzlibrary{fit}
\makeatletter
\global\let\tikz@ensure@dollar@catcode=\relax
\makeatother

\iftrack
\renewcommand{\added}[1]{{\bf #1}}
\renewcommand{\deleted}[1]{}
\renewcommand{\replaced}[2]{{\bf #2}}
\fi

\graphicspath{{./}{figures/}}

\shorttitle{Line-intensity anisotropies}
\shortauthors{Chung}
\begin{document}

\title{%How Beautifully Skewed the Sky:
A Partial Inventory of Observational Anisotropies in Single-dish Line-intensity Mapping}

%% LaTeX will automatically break titles if they run longer than
%% one line. However, you may use \\ to force a line break if
%% you desire. In v6.2 you can include a footnote in the title.

\correspondingauthor{Dongwoo T.~Chung}
\email{dongwooc@stanford.edu}

\author[0000-0003-2618-6504]{Dongwoo T.~Chung}
\affil{Physics Department and Kavli Institute for Particle Astrophysics and Cosmology, Stanford University, Stanford CA 94305, USA}

%% Mark off the abstract in the ``abstract'' environment. 
\begin{abstract}
Line-intensity mapping, being an imperfect observation of the line-intensity field in a cosmological volume, will be subject to various anisotropies introduced in observation. Existing literature in the context of CO and [\ion{C}{2}] line-intensity mapping often predicts only the real-space, spherically averaged line-intensity power spectrum, with some works considering anisotropies while examining projection of interloper emission. We explicitly consider a simplified picture of redshift-space distortions and instrumental effects due to limited resolution, and how these distort an isotropic line-intensity signal in real space and introduce strong apparent anisotropies. The results suggest that while signal loss due to limited instrumental resolution is unavoidable, measuring the quadrupole power spectrum in addition to the monopole would still break parameter degeneracies present in monopole-only constraints, even without a measurement of the full anisotropic power spectrum.
\end{abstract}

%% Keywords should appear after the \end{abstract} command. 
%% See the online documentation for the full list of available subject
%% keywords and the rules for their use.

%% From the front matter, we move on to the body of the paper.
%% Sections are demarcated by \section and \subsection, respectively.
%% Observe the use of the LaTeX \label
%% command after the \subsection to give a symbolic KEY to the
%% subsection for cross-referencing in a \ref command.
%% You can use LaTeX's \ref and \label commands to keep track of
%% cross-references to sections, equations, tables, and figures.
%% That way, if you change the order of any elements, LaTeX will
%% automatically renumber them.
%%
%% We recommend that authors also use the natbib \citep
%% and \citet commands to identify citations.  The citations are
%% tied to the reference list via symbolic KEYs. The KEY corresponds
%% to the KEY in the \bibitem in the reference list below. 

\section{Introduction}

Line-intensity mapping (or intensity mapping; LIM or IM) is the technique of observing aggregate emission in a given spectral line across a large cosmological volume, in order to make cosmological and astrophysical inferences ranging from cosmic star-formation history to dark energy and modified gravity (see~\citealt{Kovetz2017},~\citealt{Kovetz2019}, and references therein for a high-level overview). The fact that such inferences should require only observation of the aggregate emission---as opposed to individual emitters---motivates the development of small- to mid-scale dedicated instruments for line-intensity surveys, trading spatial or spectral resolution for greater survey depth. Such instruments will be complementary to large-scale interferometers and single-dish telescopes capable of community science that requires resolution of individual sources.

The design of dedicated line-intensity mappers requires a model for the expected signal, in order to set goals for instrumental sensitivities. The astronomical literature has seen a great abundance of signal forecasts, and particularly in recent years significant activity around forecasts of high-redshift ($z>1$) emission in carbon monoxide (CO) and ionised carbon ([\ion{C}{2}]) lines~\citep{Lidz11,Gong11,Gong12,Breysse2014,Uzgil14,Mashian2015,Silva2015,Yue15,Breysse2016,Li16,Serra16,Padmanabhan2018a,Padmanabhan2018b,Dumitru2019,MDK19a,MDK19}. However, much of this work considers the signal in real space rather than in redshift space, and only the spherically averaged power spectrum. Some exceptions to this generalisation are~\cite{Gong14},~\cite{LidzTaylor16}, and~\cite{Cheng16}, which all consider line-of-sight anisotropies from peculiar velocities of emitters and from erroneous projection of interloper emission to a higher-redshift comoving volume.

The aim of the present work is to extend and expand on these earlier works by explicitly considering a toy model of a set of distortions---both astrophysical and instrumental---that an observation may potentially introduce into the signal and how a line-intensity mapper may use a \emph{full} understanding of such effects to considerable advantage. Specifically, measuring the quadrupole power spectrum---and thus the leading apparent anisotropies in structure---may be sufficient to recapture information typically considered `lost' in the spherically averaged (monopole) power spectrum, and could break certain parameter degeneracies in the signal model.

The structure of the paper is as follows. In~\autoref{sec:formalism} we review and establish the mathematics around models of the signal itself and the distortions introduced in observation, before establishing power spectrum uncertainties in~\autoref{sec:uncertainties}. We then examine a specific example in~\autoref{sec:model} that demonstrates the added constraining power from considering the quadrupole power spectrum on top of the monopole, and provide a discussion of possible directions for future work in~\autoref{sec:discussion} and conclusions in~\autoref{sec:conclusions}.

Where necessary, we assume base-10 logarithms, and a $\Lambda$CDM cosmology consistent with~\cite{Planck15}: $\Omega_m = 0.307$, $\Omega_\Lambda = 0.693$, $H_0=100h$\,km\,s$^{-1}$\,Mpc$^{-1}$ with $h=0.677$, $\sigma_8 =0.816$, and $n_s =0.967$. Distances carry an implicit $h^{-1}$ dependence throughout, which propagates through masses (all based on virial masses $\propto h^{-1}$) and volume densities ($\propto h^3$).

\section{Formalism}
\label{sec:formalism}

The present work will ultimately undertake a calculation of the apparent line-intensity power spectrum and leading anisotropies, starting with the real-space signal in~\autoref{sec:realspace} and adding observational anisotropies in subsequent ones.

Numerous previous works have undertaken analytic calculations of line-intensity power spectra; the principal references for the present work are~\cite{Lidz11},~\cite{LidzTaylor16}, and~\cite{BreysseAlexandroff19}.

\subsection{Real-space Power Spectrum}
\label{sec:realspace}

Consider the line emission from sources at some fixed redshift $z$. We assume each line emitter is associated with a dark matter halo, such that a halo of virial mass $M$ has line luminosity $L(M)$ at the fixed redshift. Then we can take a halo mass function that prescribes a number density per mass bin $dn/dM$ at redshift $z$, and find the expected cosmic line-luminosity density:
\begin{equation}
    \bar\rho_L = \int dM\,\frac{dn}{dM}\,L(M).
\end{equation}
Furthermore, if halos of mass $M$ trace the underlying dark matter density fluctuations with halo bias $b(M)$, then we may obtain a luminosity-averaged bias for the emission line being considered:
\begin{equation}
    b = \frac{\int dM\,dn/dM\,L(M)\,b(M)}{\int dM\,dn/dM\,L(M)}.
\end{equation}
The luminosity contrast in this cosmic volume traces the matter density contrast with this bias $b$, such that for a given real-space matter power spectrum $P_m(k)$ as a function of comoving wavenumber $k$, the power spectrum of the luminosity contrast will be $b^2P_m(k)$. However, the observable is not the luminosity contrast or even the fluctuations in intrinsic luminosity density---which would be the contrast scaled by $\bar\rho_L$---but rather the resulting fluctuations in average surface brightness throughout the surveyed volume. The historical convention in radio astronomy is to deal with surface brightness as Rayleigh-Jeans brightness temperature; the conversion between $\bar\rho_L$ and average brightness temperature\footnote{Converting $\bar\rho_L$ into apparent spectral intensity is done just as readily: by replacing $C_{LT}$ with a factor of $c/[4\pi\nu_\text{rest}H(z)]$, one may forecast power spectra involving units of spectral brightness rather than temperature.} is given by~\cite{BreysseAlexandroff19} as
\begin{equation}
    \avg{T}=\underbrace{\left(\frac{c^3(1+z)^2}{8\pi k_B\nu_\text{rest}^3H(z)}\right)}_{C_{LT}}\bar\rho_L,
\label{eq:CLT}\end{equation}
where $\nu_\text{rest}$ is the rest frequency of the emission line, $c$ the speed of light, $k_B$ Boltzmann's constant, and $H(z)$ the Hubble parameter at redshift $z$. Then the component of the line-intensity power spectrum tracing the clustering of the underlying dark matter is given simply by $\avg{T}^2b^2P_m(k)$.

There is however a scale-independent part to the line-intensity power spectrum, the shot-noise component. Even if line emitters were completely randomly distributed without any clustering, they are discrete objects and would obey Poisson statistics. The resulting random fluctuations are described by the average squared line-luminosity density, and the shot-noise component of the power spectrum is thus
\begin{equation}
    P_\text{shot} = C_{LT}^2\int dM\,\frac{dn}{dM}\,L^2(M),
\end{equation}
where $C_{LT}$ is the same factor described in~\autoref{eq:CLT}.

The upshot is that the real-space line-intensity power spectrum is the sum of the clustering and shot-noise components:
\begin{equation}P(k) = \avg{T}^2b^2P_m(k) + P_\text{shot}.
\label{eq:Pk_realspace}\end{equation}

So far we have assumed that a halo of virial mass $M$ has the exact line luminosity given by $L(M)$, but realistically we do not expect the virial mass of a halo to be a perfect predictor of its line luminosity. Observational work and simulations suggest that baryonic physics such as stellar feedback (including feedback from supernovae), which the galaxy's halo mass can hardly indicate, lead to individual galaxies (both local and high-redshift) having bursty star-formation histories~\citep{Guo16,Feldmann17,Sparre17,Ma18,Broussard19}.

The simplest way to reflect this in our model is to prescribe an intrinsic scatter in the $L(M)$ relation. Such scatter would not affect $\avg{T}$ or $b$, both of which reflect the first moment of the $L(M)$ relation and thus average out the variability of individual halo luminosities. But scatter will affect the second moment of the $L(M)$ relation and thus $P_\text{shot}$, which as a result reflects both the average $L(M)$ relation and the level of stochasticity in line emission. We will return to scatter in~\autoref{sec:model} when calculating the $P(k)$ for a specific $L(M)$.

\subsection{Kaiser Effect Corrections}
The observed redshift of a line emitter will include not only its moving away from the observer with the Hubble flow---its cosmological redshift, so to speak---but also any peculiar velocities it may have, particularly due to the gravitational influence of nearby objects or structures. Thus, all redshift surveys---either of individual emitters or of aggregate spectral line emission---have a view of large-scale structure subjected to redshift-space distortions (RSD). \cite{Hamilton1998} provides a review of RSD in the linear regime, and outlines a line-of-sight squashing of structure at large scales---derived by~\cite{Kaiser1987}---as well as effects at smaller scales, both the `finger-of-God' effect in linear theory and damping from random velocity dispersions beyond linear theory (but with phenomenological descriptions). The formulae of~\cite{Hamilton1998} describing these effects are still largely current, applying directly to the context of halo models~\citep{Seljak2001,White2001}, and are used with minor alterations as indicated.\footnote{In addition to references in published literature, this work made use of Shun Saito's notes on redshift-space distortions, available at the time of writing at~\url{https://wwwmpa.mpa-garching.mpg.de/~komatsu/lecturenotes/Shun_Saito_on_RSD.pdf}.}

The Kaiser effect at large scales comes from the coherent motion of halos out of under-densities and into over-densities, which does not distort angular positions but results in over-densities being squashed in redshift space compared to their appearance in real space. If $P_m(k)$ is the matter power spectrum in real space, then the Kaiser effect thus amplifies the power spectrum in redshift space as the over-densities are over-emphasised.

We assume that any observational anisotropies will still preserve symmetry between the two angular dimensions---which is certainly the case for the Kaiser effect---so that even the full three-dimensional power spectrum $P(\mathbf{k})$ will depend solely on $k=|\mathbf{k}|$ and $\mu=\hat{\mathbf{k}}\cdot\hat{\mathbf{z}}$, where $\hat{\mathbf{z}}$ is the unit vector along the line of sight and thus $\mu$ the cosine of the angle between $\mathbf{k}$ and the line of sight. Then we may simply consider the anisotropic power spectrum $P(k,\mu)$ of any three-dimensional field in observation space, in lieu of $P(\mathbf{k})$.

The upshot of the Kaiser effect is that for a biased tracer of the dark matter contrast whose real-space power spectrum is given by $P(k)$, then given $f=\Omega_m(z)^\gamma$ with $\gamma\approx0.55$ for all $\Lambda$CDM models~\citep{Linder2005}, and $\beta\equiv f/b$, the redshift-space power spectrum $P^r(k,\mu)$ is given simply by
\begin{equation}
    P^r(k,\mu) = (1+\beta\mu^2)\added{^2}P(k).\label{eq:Kaiserform}
\end{equation}
Applying this to the clustering component of the line-intensity power spectrum of~\autoref{eq:Pk_realspace}, we obtain
\begin{equation}P(k,\mu) = \avg{T}^2b^2(1+\beta\mu^2)^2P_m(k)+P_\text{shot}.\label{eq:Pk_Kaiser}\end{equation}
The Kaiser effect modifies neither the Poisson statistics of the line emitters nor, consequently, the shot-noise component of the power spectrum.

As in Equation 5.5 of~\cite{Hamilton1998}, we may then expand $P(k,\mu)$ over a basis of Legendre polynomials in $\mu$:
\begin{equation}P_\ell(k)=\frac{2\ell+1}{2}\int_{-1}^1d\mu\,P(k,\mu)\mathcal{L}_\ell(\mu),\label{eq:legendre}\end{equation}
where $\mathcal{L}_\ell$ denotes the Legendre polynomial of order $\ell$. Then
\begin{equation}
    P(k,\mu)=\sum_\ell P_\ell(k)\mathcal{L}_\ell(\mu).
\end{equation}
Since $P(k,\mu)$ of~\autoref{eq:Pk_Kaiser} only has terms of order $\mu^0$, $\mu^2$, and $\mu^4$, $P_\ell(k)$ can only be non-vanishing for $\ell=0$, $\ell=2$, and $\ell=4$, at this level of consideration. The observables that we propose are the monopole power spectrum $P_{\ell=0}(k)$, which is the best redshift-space counterpart to the conventionally considered spherically symmetric $P(k)$, and the quadrupole $P_{\ell=2}(k)$. While the hexadecapole $P_{\ell=4}(k)$ is also non-zero, we omit detailed discussion of it from the main text; the discussion of~\autoref{sec:model} will show that considering anisotropies beyond the quadrupole does not appear to drastically improve the constraining power of the observation. For completeness, however,~\autoref{sec:hexadecapole} calculates counterparts to most monopole- and quadrupole-related quantities we discuss in the remainder of the main text.

If we simply substitute $P(k,\mu)$ of~\autoref{eq:Pk_Kaiser} and the appropriate Legendre polynomials into~\autoref{eq:legendre}, we obtain
\begin{align}P_{\ell=0}(k)&=\left(1+\frac{2}{3}\beta+\frac{1}{5}\beta^2\right)\avg{T}^2b^2P_m(k)+P_\text{shot};\label{eq:Kaiser1}\\
P_{\ell=2}(k)&=\left(\frac{4}{3}\beta+\frac{4}{7}\beta^2\right)\avg{T}^2b^2P_m(k).\label{eq:Kaiser2}\end{align}
As the shot noise does not depend on $\mu$, it would not appear to contribute to the quadrupole. Thus, at this point one might suppose that the quadrupole power spectrum is a pure measurement of the clustering of the line emission.

\subsection{Observational Anisotropies from Instrument Resolution}

Previous literature certainly accounts for the smearing of line-intensity fluctuations from limited angular and spectral resolution, but folds this smearing into the uncertainties rather than the observed signal. However, \replaced{since the instrumental response can introduce significant anisotropies into the observation, we should}{just as we consider redshift-space distortions above as a modification of the signal rather than a factor in the uncertainties, here we will} consider the signal after convolution with this response. This approach\added{ leads to equivalent detection significances and errors on inferred parameters, but} may lead to different insights\added{ about what is or is not observable in a single-dish survey,} compared to considering the response as part of the uncertainties.% For instance, if an instrument's native spectral resolution far exceeds the science channelisation, the signal will not smear along the line of sight to any noticeable extent. The beam of the instrument, however, will unavoidably smooth the angular map in each frequency channel.

Suppose we survey the three-dimensional line-intensity field $T(\mathbf{x})$ (again, as brightness temperature) with an instrument of finite angular and spectral resolution. Assume the beam profile is approximately Gaussian with standard deviation $\sigma_\text{beam}$ (in radians), and the frequency response is also Gaussian with standard deviation $\sigma_\nu$. We refer to Section C.3 of~\cite{Li16} to see how this affects the signal in apparent comoving space.

If the map is at redshift $z$ at a comoving distance $R(z)$ from the observer, then the beam profile corresponds to a Gaussian profile in comoving space with standard deviation
\begin{equation}
    \sigma_\perp=R(z)\sigma_\text{beam},
    \label{eq:sigmaperp}
\end{equation}
and the spectral response to a Gaussian profile in comoving space with standard deviation
\begin{equation}
    \sigma_\parallel=\frac{c(1+z)^2}{H(z)}\frac{\sigma_\nu}{\nu_\text{rest}}.
    \label{eq:sigmaparallel}
\end{equation} The Fourier transform of a normalised Gaussian profile $G(x_i)=(2\pi\sigma_i)^{-1/2}\exp{[-x_i^2/(2\sigma_i)^2]}$ is $\tilde{G}(k_i)=\exp{(-k_i^2\sigma_i^2/2)}$, and convolution with this profile in each of the dimensions in real space becomes multiplication in Fourier space:
\begin{align}\tilde{T}_\text{conv}(\mathbf{k})&=\tilde{T}(\mathbf{k})\tilde{G}(k_1)\tilde{G}(k_2)\tilde{G}(k_3)\notag\\&=\tilde{T}(\mathbf{k})\exp{(-k_\perp^2\sigma_\perp^2/2-k_\parallel^2\sigma_\parallel^2/2)},\label{eq:Tconv}\end{align}
where $k_3=k_\parallel=k\mu$ is the component of $\mathbf{k}$ along the line of sight, while the first and second dimensions are in the transverse directions, and thus $k_\perp^2=k_1^2+k_2^2=k\sqrt{1-\mu^2}$. Since the power spectrum $P(\mathbf{k})$ may be found as $|\tilde{T}(\mathbf{k})|^2/V_\text{surv}$,
\begin{align}P_\text{conv}(\mathbf{k})&=P(\mathbf{k})\exp{(-k_\perp^2\sigma_\perp^2-k_\parallel^2\sigma_\parallel^2)}\\&=P(\mathbf{k})\exp{[-k^2\sigma_\perp^2(1-\mu^2)-k^2\sigma_\parallel^2\mu^2]}.\end{align}
Once compressed into the two-dimensional space of $(k,\mu)$, the effect of convolution with the instrument response should persist:
\begin{align}P_\text{conv}(k,\mu)&=P(k,\mu)\exp{[-k^2\sigma_\perp^2(1-\mu^2)-k^2\sigma_\parallel^2\mu^2]}\\&=[\avg{T}^2b^2(1+\beta\mu^2)^2P_m(k)+P_\text{shot}]\notag\\*&\hspace{1cm}\times\exp{[-k^2\sigma_\perp^2(1-\mu^2)-k^2\sigma_\parallel^2\mu^2]}.\end{align}
To see what monopole and quadrupole power spectra we could realistically expect to observe, then, we should not use~\autoref{eq:Kaiser1} and~\autoref{eq:Kaiser2}, but rather substitute this expression into the integral by which we obtain $P_\ell(k)$. Then it is no longer obvious that $P_\text{shot}$ will disappear from the quadrupole, given that it is being convolved with what is potentially a highly anisotropic instrument response.

Define $C_0$, $S_0$, $C_2$, and $S_2$ such that
\begin{align}P_{\text{conv},\ell=0}(k)&=C_0\avg{T}^2b^2P_m(k)+S_0P_\text{shot};\label{eq:Kaiserconv1}\\
P_{\text{conv},\ell=2}(k)&=C_2\avg{T}^2b^2P_m(k)+S_2P_\text{shot}.\label{eq:Kaiserconv2}\end{align}
For brevity, define also $\alpha_\perp = k\sigma_\perp$, $\alpha_\parallel = k\sigma_\parallel$, and $\delta_\alpha^2=\alpha_\parallel^2-\alpha_\perp^2$. Via Mathematica\footnote{Wolfram Research, Inc.; Version 11.2, 2017.}, we obtain
\begin{align}
    C_0&= \frac{1}{8}\left\{\frac{\pi^{1/2}\exp{(-\alpha_\perp^2)}\erf{[(\delta_\alpha^2)^{1/2}]}}{(\delta_\alpha^2)^{5/2}}\left[4\delta_\alpha^4+4\delta_\alpha^2\beta+3\beta^2\right]\right.\notag\\*&\left.\hspace{2cm}-\frac{2\beta\exp{(-\alpha_\parallel^2)}}{\delta_\alpha^4}[2\delta_\alpha^2(\beta+2)+3\beta]\right\};\label{eq:Kaiserconv3}\\
    S_0&= \frac{\pi^{1/2}\exp{(-\alpha_\perp^2)}\erf{[(\delta_\alpha^2)^{1/2}]}}{2(\delta_\alpha^2)^{1/2}};\label{eq:Kaiserconv4}\\
    C_2&= \frac{5}{32}\left\{\frac{\pi^{1/2}\exp{(-\alpha_\perp^2)}\erf{[(\delta_\alpha^2)^{1/2}]}}{(\delta_\alpha^2)^{7/2}}\left[45\beta^2-6\delta_\alpha^2(\beta^2-6\beta)\right.\right.\notag\\*&\hspace{4cm}\left.\left.+4\delta_\alpha^4(3-2\beta)-8\delta_\alpha^6\right]\right.\notag\\*&\hspace{2cm}\left.-\frac{2\exp{(-\alpha_\parallel^2)}}{\delta_\alpha^6}\left[4\delta_\alpha^4(2\beta^2+4\beta+3)\right.\right.\notag\\*&\hspace{4cm}\left.\left.+12\delta_\alpha^2(2\beta^2+3\beta)+45\beta^2\right]\right\};\label{eq:Kaiserconv5}\\
    S_2&= \frac{5}{8}\left[\frac{\pi^{1/2}\exp{(-\alpha_\perp^2)}\erf{[(\delta_\alpha^2)^{1/2}]}}{(\delta_\alpha^2)^{3/2}}(3-2\delta_\alpha^2)\right.\notag\\&\left.\hspace{5cm}-\frac{6\exp{(-\alpha_\parallel^2)}}{\delta_\alpha^2}\right].
    \label{eq:Kaiserconv6}
\end{align}
(Note that if $\sigma_\parallel<\sigma_\perp$ and thus $\delta_\alpha^2<0$, all coefficients are still real, owing to the properties of the error function.)

In the limit of $\alpha_\perp\to0$ and $\alpha_\parallel\to0$, these all reduce to the limits of $C_0\to1+2\beta/3+\beta^2/5$, $S_0\to1$, $C_2\to4\beta/3+4\beta^2/7$, and $S_2\to0$---all expected, given~\autoref{eq:Kaiser1} and~\autoref{eq:Kaiser2}. However, for finite resolution and correspondingly sufficiently large values of $k$, these coefficients will shift significantly away from those limits.

\subsubsection{Examples of Instrument Resolution}
\label{sec:examplesigmas}

\begin{figure}[t!]
 \centering\includegraphics[width=\columnwidth]{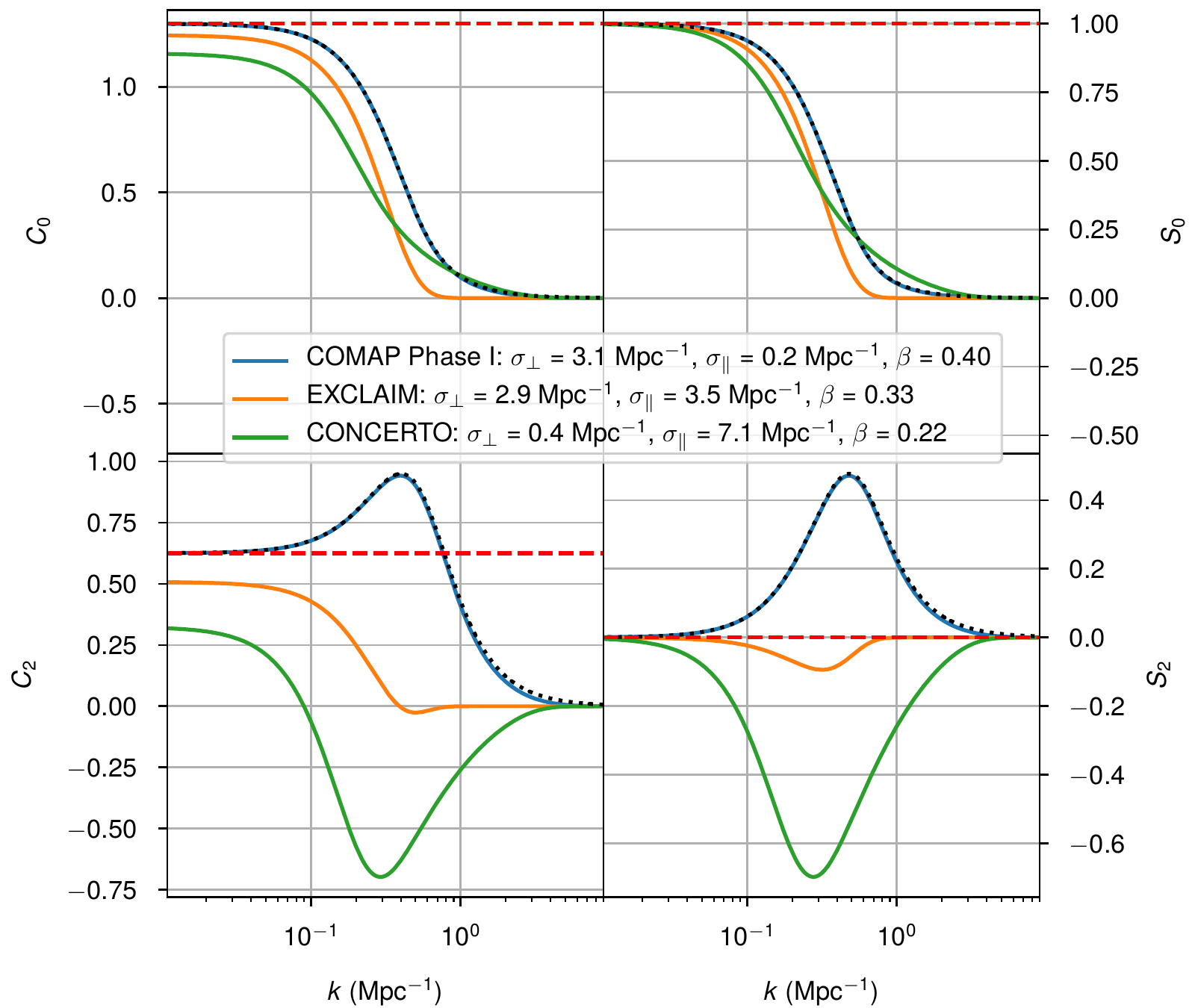}
 \caption{The coefficients $C_0$, $S_0$, $C_2$, and $S_2$ of Equations~\ref{eq:Kaiserconv1} through~\ref{eq:Kaiserconv6} as functions of $k$, given the example values of $\sigma_\perp$ and $\sigma_\parallel$ for the surveys of~\autoref{tab:examplesigmas}, and illustrative values of $\beta$ loosely based on~\cite{MDK19}. The value of $\sigma_\parallel$ in COMAP Phase I has been chosen to match $1/4$ of the nominal channel width, to reflect the higher intrinsic frequency resolution of the spectrometer. Black dotted lines in each panel show the limit of the coefficients for COMAP Phase I as $\sigma_\parallel\to0$. Red dashed lines in each panel show the limit given $\beta=0.4$ as $\sigma_\perp\to0$ and $\sigma_\parallel\to0$.}
 \label{fig:Pconv_coeffs}
\end{figure}

\begin{deluxetable*}{ccccccccc}
\tabletypesize{\footnotesize}
\tablewidth{0.9\linewidth}
\tablecaption{\label{tab:examplesigmas}
Examples of $\sigma_\parallel$ and $\sigma_\perp$ for selected line-intensity surveys.}
\tablehead{
\colhead{Survey}&\colhead{Reference}&\colhead{Emission line}&\colhead{$\nu_\text{rest}$ (GHz)}&\colhead{$z$}&\colhead{Beam width (arcmin)}&\colhead{$\delta_\nu$ (GHz)}&\colhead{$\sigma_\perp$ (Mpc)}&\colhead{$\sigma_\parallel$ (Mpc)}}
\startdata
COMAP Phase I&\cite{Ihle19}&CO(1-0)&115.27&$2.8$&4&$0.0156$&3.1&$<0.9$\\
EXCLAIM&\cite{Padmanabhan2018b}&[\ion{C}{2}]&1901&$3.0$&3.6&1&2.9&3.5\\
CONCERTO&\cite{Dumitru2019}&[\ion{C}{2}]&1901&$6.0$&0.39&1.5&0.4&7.1
\enddata
\tablecomments{The beam width is taken to be the full width at half-maximum (FWHM), which is $2.355\sigma_\text{beam}$. If the reference specifies only the dish size, we assume a diffraction-limited beam width given by 1.22 times the observation wavelength divided by the dish diameter. The quantity $\delta_\nu$ denotes the bandwidth of a spectrometer channel. For any quoted resolving power $R$, we assume the corresponding frequency channel width is $\delta_\nu=\nu/R$. We further assume $\sigma_\nu=\delta_\nu/2.355$, with $\delta_\nu$ thus taken to represent the FWHM of the frequency response profile. As~\cite{Ihle19} note, operation of the COMAP instrument with increased spectral resolution is possible, meaning that the native resolution of the COMAP spectrometer is significantly finer than the quoted $\delta_\nu$.}
\end{deluxetable*}
To demonstrate how these coefficients change with instrument response, we take three line-intensity surveys currently under development---Phase I of the Carbon mOnoxide Mapping Array Pathfinder (COMAP), targeting CO(1-0) emission at $z\sim2.8$; the EXperiment for Cryogenic Large-Aperture Intensity Mapping (EXCLAIM), targeting [\ion{C}{2}] emission at $z\sim3$; and the CarbON [\ion{C}{2}] line in post-rEionisation and Reionisation epoch project (CONCERTO), also targeting [\ion{C}{2}] emission, but at $z\sim6$). We calculate values of $\sigma_\parallel$ and $\sigma_\perp$ for each of these, as listed in~\autoref{tab:examplesigmas}, and the values are then used to calculate example values of $C_0$, $S_0$, $C_2$, and $S_2$ in~\autoref{fig:Pconv_coeffs}. Note that for illustrative purposes, we assume $\beta=0.4$ for CO(1-0) at $z\sim2.8$, $\beta=0.33$ for [\ion{C}{2}] at $z\sim3$, and $\beta=0.22$ for [\ion{C}{2}] at $z\sim6$, based on the line biases from the models in~\cite{MDK19}, and the relatively safe assumption that $\Omega_m$ is within a few percent of 1 at these redshifts. Note also that we use a smaller value of $\sigma_\parallel$ than might be inferred the nominal $15.6$ MHz value for the frequency channel width of COMAP Phase I. This is because, as~\cite{Ihle19} note, the COMAP spectrometer is capable of operating at higher spectral resolution, which means its native resolution is finer than the nominal value. In fact, the raw spectra before data reduction are in 4096 channels across 8 GHz, so the basic resolution of the spectrometer is as fine as 1.95 MHz, and eight such channels co-added result in the 15.6 MHz bandwidth per channel in~\cite{Ihle19}. We conservatively use a value of $\sigma_\parallel$ corresponding to $1/4$ of the nominal $15.6$ MHz bandwidth.

Previous literature has already anticipated attenuation of $P(k)$ at high $k$ due to limited resolution, which is what $S_0$ reflects; $C_0$ merely has a small correction that accounts for enhancement of the monopole due to the Kaiser effect. However, what previous literature does not appear to explicitly consider is the effect of limited resolution on the quadrupole, which is what $C_2$ and $S_2$ describe. These suggest that not only could there be a slight enhancement of the clustering component over a certain range of $k$, but also the shot noise will actually appear in $P_{\text{conv},\ell=2}(k)$ over a similar range. For an experiment like COMAP Phase I (which we will simply call COMAP for the remainder of the present work), with high spectral resolution but low angular resolution, this manifests as a positive addition to the quadrupole at $k\sim0.5$ Mpc$^{-1}$. For an experiment like CONCERTO, with high angular resolution but low spectral resolution, the clustering component of the quadrupole crosses zero at $k\sim0.1$ Mpc$^{-1}$ and becomes negative but enhanced at $k\sim0.3$ Mpc$^{-1}$, joined by a significant negative shot-noise component.

We pause here to emphasise a curious implication of this result alone\replaced{. Due to the emphasis on the monopole $P_{\ell=0}(k)$, the auto shot noise has traditionally been considered a high-$k$ signal, as the clustering component dominates at low $k$. For radio observations, this categorisation would put shot noise firmly in the niche of interferometers, and out of reach of single-dish instruments. However, the instrumental response is capable of mixing shot noise with clustering in the quadrupole power spectrum at intermediate ranges of $k$, at a different ratio than in the monopole. This means that single-dish instruments with highly anisotropic instrumental response---if that response is properly characterised---may yet retain some sensitivity to the line-intensity shot noise. Note, finally, that the observational anisotropy imposed on the signal would not be as obvious if the instrument response were folded into the uncertainties rather than the signal itself.}{, which is non-zero sensitivity of single-dish surveys to the line-intensity auto shot noise. To be sure, interferometers are capable of significantly higher angular resolution at these wavelengths than single-dish instruments, as it is easier to synthesise large apertures with long baselines than it is to build larger dishes. Thus, interferometers have superior sensitivity to the power spectrum at high $k$, in the regime where shot noise dominates over the clustering component of $P_{\ell=0}(k)$. The interferometric COPSS survey~\citep{COPSS}, for instance, achieves its lowest errors on $P(k)$ at $k\sim1$ Mpc$^{-1}$, a range of $k$ where the instrumental response of our example single-dish surveys attenuates the monopole power spectrum by an order of magnitude, if not to essentially zero. But on the other hand, we also explicitly find that the shot-noise component should contribute to the observed quadrupole (and at a different proportion to the clustering component than in the monopole, which should reduce degeneracies in inference), since single-dish spectrometers are still sensitive to the high-$k$ spectral (high-$|\mu|$) fluctuations, which then appear highly anisotropic from being smeared by the telescope beam.}

An implicit assumption in the above calculations is that the entire range of $\mu$ is available at each $k$ to sum or integrate over, which is necessary for a faithful estimation of $P_{\ell=2,\text{conv}}(k)$. However, as discussed in Appendix A of~\cite{UM_CII}, a finite frequency resolution $\delta_\nu$ in the data cube imposes a cutoff of $\pi$ divided by the comoving length corresponding to $\delta_\nu$:
\begin{equation}
    k_{\delta_\nu}=\pi\left[\frac{c(1+z)^2}{H(z)}\frac{\delta_\nu}{\nu_\text{rest}}\right]^{-1}.
\end{equation}
Above this $k$, the Fourier space accessible with the data cube is truncated. In the case of CONCERTO, which uses a Fourier-transform spectrometer, if there is only one resolution element per $\delta_\nu$ of bandwidth per instrumental line-of-sight, there would be no line-of-sight modes for $k>k_{\delta_\nu}=0.19$ Mpc$^{-1}$ available for analysis to measure line-of-sight anisotropies. Grating spectrometers and scanning spectrometers should be able to oversample the spectral response with appropriate design---in the case of EXCLAIM, if the spectral response is Nyquist-sampled so that the data is in frequency bins of $\delta_\nu/2=0.5$ GHz, then in fact the cutoff is not $k_{\delta_\nu}=0.38$ Mpc$^{-1}$, but twice this at $0.76$ Mpc$^{-1}$.

Overall, it is typically more sensible to build a medium-resolution ($\nu/\delta_\nu\sim10^3$) spectrometer and design a survey strategy to oversample the instrumental angular resolution, than to build a large telescope for finer angular resolution but have only a low-resolution ($\nu/\delta_\nu\lesssim10^2$) spectrometer on board. The line-of-sight smearing in the latter case subtracts from the quadrupole and therefore its detectability. Achieving such high resolving power may be fundamentally more of a challenge in some experimental contexts than others---millimetre-wave direct-detection spectrometers like EXCLAIM and CONCERTO, for instance, have somewhat different sensitivity considerations compared to centimetre-wave radiometers like COMAP. As such, some line-intensity surveys may only be able to probe these anisotropies by targeting cross-correlation with a different tracer covering the same survey volume, which may allow sacrificing sensitivity to the line-intensity auto-correlation to gain resolving power.

\subsubsection{Coefficient Limits as $\sigma_\parallel\to0$}

Given the caveat that the instrument response must be over-sampled during the observation, we focus on the case where $\sigma_\parallel\ll\sigma_\perp$, since oversampling the sky is a more natural and potentially more economical thing to do than over-allocating detectors in an array for finer sampling of the spectrometer channel frequency response. In the limit of $\sigma_\parallel\to0$, the coefficients of Equations~\ref{eq:Kaiserconv1} through~\ref{eq:Kaiserconv6} become
\begin{align}
    C_0&=\frac{\alpha_\perp\beta[2\alpha_\perp^2(\beta+2)-3\beta]+(4\alpha_\perp^4-4\alpha_\perp^2\beta+3\beta^2)F(\alpha_\perp)}{4\alpha_\perp^5};\label{eq:Kaiserconv3_}\\
    S_0&=F(\alpha_\perp)/\alpha_\perp;\label{eq:Kaiserconv4_}\\
    C_2&=\frac{5}{16\alpha_\perp^7}\left\{4\alpha_\perp^5(2\beta^2+4\beta+3)-12\alpha_\perp^3\beta(2\beta+3)+45\alpha_\perp\beta^2\right.\notag\\*&\left.\hspace{2cm} -[8\alpha_\perp^6+4\alpha_\perp^4(3-2\beta)+6\alpha_\perp^2(\beta^2-6\beta)\right.\notag\\*&\left.\hspace{4cm}+45\beta^2]F(\alpha_\perp)\right\};\label{eq:Kaiserconv5_}\\
    S_2&=\frac{5}{4\alpha_\perp^3}[3\alpha_\perp-(2\alpha_\perp^2+3)F(\alpha_\perp)].\label{eq:Kaiserconv6_}
\end{align}
Here, $F(x)$ denotes Dawson's integral (as considered by~\citealt{Rybicki89}). Based on the results of~\autoref{fig:Pconv_coeffs}, COMAP approaches this limit closely enough that we will focus on it and use these simpler formulae for the remainder of this work.

\subsection{Small-scale Corrections}
A small-scale correction to the power spectrum applies due to the damping of apparent structure from random peculiar velocities within collapsed objects. \cite{Hamilton1998} describes either a Gaussian damping based on a Gaussian pairwise velocity distribution or a Lorentzian damping based on an exponential distribution; the present work will consider the Gaussian damping function used by~\cite{Seljak2001} and~\cite{White2001}, as well as by~\cite{Cheng16} in a line-intensity mapping context. To apply this damping, we multiply the clustering component of our power spectrum by a factor of $\exp{(-k^2\mu^2\sigma_p^2/2)}$:
\begin{align}
    P_\text{obs}(k,\mu)&=[\avg{T}^2b^2(1+\beta\mu^2)^2\exp{(-k^2\mu^2\sigma_p^2/2)}P_m(k)+P_\text{shot}]\notag\\*&\hspace{1cm}\times\exp{[-k^2\sigma_\perp^2(1-\mu^2)-k^2\sigma_\parallel^2\mu^2]}.
    \label{eq:Pk_damped}
\end{align}
Here, $\sigma_p$
corresponds to a pairwise velocity dispersion (divided by \replaced{$H(z)$}{$aH(z)=H(z)/(1+z)$} to make it a comoving scale). For the COMAP signal, we assume $\sigma_p=\replaced{300}{70}$ km s$^{-1}$ for the dispersion (equivalent at $z=2.8$ to a scale of \replaced{1}{0.9} Mpc), which is not unreasonable given the\added{ fit} value from~\cite{Taruya2010} of the single-point velocity dispersion (which is $2^{1/2}$ times smaller than the pairwise dispersion) near \replaced{200}{50} km s$^{-1}$ at $z\sim3$\added{ (given a Gaussian damping function and the linear matter power spectrum, the same assumptions as in the present work)}. We will not re-calculate the results of Equations~\ref{eq:Kaiserconv3_} through~\ref{eq:Kaiserconv6_} with the damping term, and numerically evaluate the resulting total observed power spectrum when taking this damping into account.

It is also possible to consider a one-halo component of the power spectrum, based on pairwise correlations between line emitters hosted by a single halo, and this will also have redshift-space distortions. This one-halo term may be considered in future work, but is beyond the scope of the present work, which assigns only one line emitter to a halo.

\subsection{Summary: Everything But the Projection Effect}
We briefly recap all observational anisotropies considered:
\begin{itemize}
    \item corrections using the Kaiser formula, due to large-scale coherent infall into overdensities squashing apparent structures along the line of sight;
    \item limited instrumental resolution smearing structures along either transverse or line-of-sight dimensions;
    \item and `finger-of-God' damping at high $k$ due to random pairwise velocity dispersions smearing small-scale structure.
\end{itemize}
Since we are considering only the signal of interest, we do not consider the projection effect that applies to interloper emission; this is beyond the scope of the present work, and is already well covered by~\cite{Gong14},~\cite{LidzTaylor16}, and~\cite{Cheng16}.

We also present in~\autoref{tab:cartoons} schematic illustrations of how each effect listed above distorts a spherical overdensity in observation. If spherical overdensities look squashed, the effect adds to the quadrupole; if spherical overdensities look elongated, the effect subtracts from the quadrupole.

\begin{deluxetable*}{m{1.5cm}ccc}
\tabletypesize{\footnotesize}
\tablewidth{0.9\linewidth}
\tablecaption{\label{tab:cartoons}
Effect of different observational anisotropies on the appearance of a spherical matter overdensity.}
\tablehead{
\colhead{Effect}&\colhead{Direction of squashing/smearing}&\colhead{Root cause}&\colhead{Additive/subtractive quadrupole?}}
\startdata
Kaiser effect (linear RSD)&
    \begin{tikzpicture}[baseline=0]
    \draw (0,0) circle (0.5);
    \draw[->] (0,0.5) -- (0,0.25);
    \draw[->] (0,-0.5) -- (0,-0.25);
    \draw[->] (0.5,0) -- (0.25,0);
    \draw[->] (-0.5,0) -- (-0.25,0);
    \draw[color=red] (0,0) ellipse (0.5 and 0.25);
    \node[fit=(current bounding box),inner ysep=1mm,inner xsep=0]{}; %thanks https://tex.stackexchange.com/questions/152067
    \end{tikzpicture}
    &Coherent infall of sources into overdensity&Additive\\
Instrumental resolution: $\sigma_\perp\gg\sigma_\parallel$&
    \begin{tikzpicture}[baseline=0]
    \draw (0,0) circle (0.5);
    \draw[->,color=gray] (0,0.5) -- (0,0.6);
    \draw[->,color=gray] (0,-0.5) -- (0,-0.6);
    \draw[->,color=gray] (0.5,0) -- (1.25,0);
    \draw[->,color=gray] (-0.5,0) -- (-1.25,0);
    \draw[color=red] (0,0) ellipse (1.25 and 0.6);
    \node[fit=(current bounding box),inner ysep=1mm,inner xsep=0]{};
    \end{tikzpicture}
    &Transverse smearing from limited angular resolution&Additive\\
Instrumental resolution: $\sigma_\parallel\gg\sigma_\perp$&
    \begin{tikzpicture}[baseline=0]
    \draw (0,0) circle (0.5);
    \draw[->,color=gray] (0.5,0) -- (0.6,0);
    \draw[->,color=gray] (-0.5,0) -- (-0.6,0);
    \draw[->,color=gray] (0,0.5) -- (0,1.25);
    \draw[->,color=gray] (0,-0.5) -- (0,-1.25);
    \draw[color=red] (0,0) ellipse (0.6 and 1.25);
    \node[fit=(current bounding box),inner ysep=1mm,inner xsep=0]{};
    \end{tikzpicture}
    &Line-of-sight smearing from limited spectral resolution&Subtractive\\
`Finger-of-God' damping&
    \begin{tikzpicture}[baseline=0]
    \draw (0,0) circle (0.5);
    \draw[->] (0.3,-0.4) -- (-0.9,0.6);
    \draw[->] (-0.4,0.3) -- (0.9,0);
    \draw[->] (0,0.5) -- (0.1,-1.3);
    \draw[->] (0,-0.5) -- (-0.3,1.25);
    \draw[color=red] (0,0) ellipse (0.5 and 1.25);
    \node[fit=(current bounding box),inner ysep=1mm,inner xsep=0]{};
    \end{tikzpicture}
    &Random velocities within (and infall through) collapsed overdensity&Subtractive\\
\enddata
\tablecomments{The observer line-of-sight runs up the page. The red ellipses indicate the distorted appearance of the original spherical overdensity (black circles). Black arrows indicate actual motions of sources within the overdensity, while grey arrows indicate instrumental smearing.}
\end{deluxetable*}

\section{Power Spectrum Uncertainties}
\label{sec:uncertainties}
In~\autoref{sec:model}, we will take the observables calculated up to this point and consider how detectable they are and what constraints they can provide. To do this, we must consider uncertainties on power spectra, and in turn we must first consider how the spectra are calculated. From a three-dimensional map of the line-intensity field on a discrete grid of comoving positions, we can calculate a three-dimensional power spectrum $P(\mathbf{k})$, also on a discrete grid (of comoving wavenumbers). The most straightforward, na\"{i}ve estimator of the spherically symmetric power spectrum is then
\begin{equation}P(k)=\frac{\sum_{\mathbf{k}:|(|\mathbf{k}|-k)|<\Delta k}P(\mathbf{k})}{\sum_{\mathbf{k}:|(|\mathbf{k}|-k)|<\Delta k}1},\end{equation}
an average over modes in $k$-shells of width $\Delta k$. We may define the number of modes $N_m(k)$ as the denominator of the above equation.

Similarly, we can estimate the anisotropic power spectrum as
\begin{equation}P(k,\mu)=\frac{\sum_{\mathbf{k}:|(|\mathbf{k}|-k)|<\Delta k\text,|(k_\parallel/|\mathbf{k}|-\mu)|<\Delta \mu}P(\mathbf{k})}{\sum_{\mathbf{k}:|(|\mathbf{k}|-k)|<\Delta k,|(k_\parallel/|\mathbf{k}|-\mu)|<\Delta \mu}1},\end{equation}
\replaced{which now averages}{averaging} over $\mu$-sectors of width $\Delta\mu$ within the previously defined $k$-shells. We define $N_m(k,\mu)$ as the denominator of this equation.\footnote{Unless explicitly otherwise indicated, $N_m(k)$ refers to the number of modes in the whole $k$-shell, and $N_m(k,\mu)$ to the number of modes in the intersection of the $k$-shell and $\mu$-sector.}

In the present work, we will not use power spectra derived from mock data cubes of line-intensity fields, and need analytic expressions for $N_m(k)$ and $N_m(k,\mu)$. If we consider linear $k$-bins of width $\Delta k$, and linear $\mu$-bins of width $\Delta \mu$, then Equation 21 of~\cite{LidzTaylor16} becomes
\begin{equation}
    N_m(k,\mu) = \frac{k^2V_\text{surv}}{4\pi^2}\Delta k\Delta\mu,
    \label{eq:Nmkmu}
\end{equation}
where $V_\text{surv}$ is the total comoving volume surveyed\added{ within the targeted patch area}. Throughout this work, we will only deal with $\mu\in(0,1)$, as only half the Fourier modes are independent in the power spectrum of a real-valued field. Then, integrating the above over $\mu\in(0,1)$ yields simply
\begin{equation}
    N_m(k) = \frac{k^2V_\text{surv}}{4\pi^2}\Delta k.
    \label{eq:Nmk}
\end{equation}

Now consider the uncertainties. For this work, we consider only thermal instrumental noise and sample variance, discarding any additional uncertainties from systematics or data cleaning. This still means that in addition to the signal $P(\mathbf{k})$, there is added noise in the total observed $P_\text{total}(\mathbf{k})$. \replaced{T}{For a radiometer like COMAP\footnote{The formalism around noise differs for experiments like CONCERTO and EXCLAIM, which operate at higher frequency with non-radiometer architectures; see~\cite{UM_CII} and~\cite{Padmanabhan2018b}.}, prior works like~\cite{Li16} show that t}he thermal noise contributes a completely $k$- and $\mu$-independent component with an expectation value of
\begin{equation}
    P_n = \frac{T_\text{sys}^2}{\delta_\nu t_\text{pix}}V_\text{vox} = \frac{T_\text{sys}^2}{\delta_\nu}\frac{(\Omega_\text{surv}/\Omega_\text{pix})}{N_\text{feeds}t_\text{surv}}V_\text{vox},
\end{equation}
where $T_\text{sys}$ is the system temperature, $\delta_\nu$ is (again) the bandwidth per frequency channel, $t_\text{pix}$ is the observation time per map pixel, and $V_\text{vox}$ is the comoving volume per voxel (or, per map pixel per $\delta_\nu$). The above equation then expands $t_\text{pix}$ into an expression in terms of the number of feeds $N_\text{feeds}$, the total on-sky observation time $t_\text{surv}$, the total survey patch solid angle $\Omega_\text{surv}$, and the map pixel solid angle $\Omega_\text{pix}$. Since (by analogy to~\autoref{eq:sigmaperp} and~\autoref{eq:sigmaparallel})
\begin{equation}V_\text{vox}=R^2(z)\Omega_\text{pix}\cdot\frac{c(1+z)^2}{H(z)}\frac{\delta_\nu}{\nu_\text{rest}},\end{equation}
the noise power spectrum is ultimately independent of the voxel parameters:
\begin{equation}
    P_n = \frac{T_\text{sys}^2\Omega_\text{surv}}{N_\text{feeds}t_\text{surv}}\frac{c(1+z)^2R^2(z)}{H(z)\nu_\text{rest}}.
    \label{eq:Pn}
\end{equation}

For the monopole $P_{\ell=0}(k)$, the error is well-established (in~\citealt{Lidz11}, for instance):
\begin{equation}\sigma_{P_{\ell=0}}(k) = \frac{P_{\ell=0}(k)+P_n}{\sqrt{N_m(k)}},\end{equation}
where we may consider $P(k)/\sqrt{N_m(k)}$ to be the contribution to the uncertainty from sample variance, and $P_n/\sqrt{N_m(k)}$ the contribution from thermal noise. Note in particular that since the data cube is of finite extent, we will not see a perfectly flat contribution of $P_n$ to all $P_\text{total}(\mathbf{k})$, but a random contribution of mean $P_n$ and standard deviation $P_n$.

We may apply this straightforwardly to $P(k,\mu)$, as~\cite{LidzTaylor16} and~\cite{Cheng16} do:
\begin{equation}\sigma_P(k,\mu) = \frac{P(k,\mu)+P_n}{\sqrt{N_m(k,\mu)}}.\end{equation}
Note that in all cases, the signal power spectrum included in the expressions for $\sigma_P$ is not the real-space or even redshift-space power spectrum, but the power spectrum of, say,~\autoref{eq:Pk_damped}, incorporating \emph{all} observational anisotropies.

For the quadrupole, no existing literature appears to explicitly derive the uncertainties\added{ for line-intensity mapping}. While a full derivation of the covariance matrix for $P_{\ell=0}(k)$ and $P_{\ell=2}(k)$ is beyond the scope of this work, we \replaced{consider a scenario where uncertainties on the quadrupole measurement are dominated by thermal noise (which also avoids significant off-diagonal covariances). Re-writing~\autoref{eq:legendre} in our specific case of $\ell=2$:
\begin{equation}P_{\ell=2}(k)=\frac{5}{2}\int_0^1d\mu\,P(k,\mu)(3\mu^2-1),\end{equation}
where we have assumed the integrand is even in $\mu$.

Now consider this calculation over a discrete data cube of nothing more than Gaussian noise, so that the resulting $P(\mathbf{k})$ is randomly distributed with mean $P_n$ and standard deviation $P_n$. Associated with each of these is an uncorrelated value of $|\mu|$ between $0$ and 1. If we sample from the grid in $\mathbf{k}$-space a random value of $\mathbf{k}$, we may consider it to have a random value of $P(\mathbf{k})$ and a random value of $\mu^2$. The specific details of the distributions of each does not matter---since we will end up averaging over $N_m(k)$ anyway, all we care about is the variance. We already know $\sigma^2[P(\mathbf{k})]=P_n^2$, and it is straightforward to calculate that
\begin{equation}\sigma^2[\mu^2]=\avg{\mu^4}-\avg{\mu^2}^2=4/45.\end{equation}
Then we may calculate\footnote{Note that in general, there is an additional term in finding the variance of the product of two random variables, but in~\autoref{eq:42} this would be $\sigma^2[P(k)]\avg{3\mu^2-1}^2$, which evaluates to zero.}
\begin{align}
    \sigma^2[(3\mu^2-1)P(\mathbf{k})] &= \sigma^2[3\mu^2-1]\left\{\sigma^2[P(\mathbf{k})]+\avg{P(\mathbf{k})}^2\right\}\label{eq:42}\\&=\frac{8}{5}P_n^2,
\end{align}
Applying the appropriate factor of $(5/2)^2$ to convert this to a variance on $P_{\ell=2}(k)$ ultimately leads us to the expression}{adapt Equation C1 of~\cite{Taruya2010}, with galaxy power spectrum multipoles and homogeneous shot noise replaced by line-intensity power spectrum multipoles and homogeneous instrumental thermal noise:
\begin{align}\operatorname{cov}{[P_\ell(k),P_{\ell'}(k)]}&=\frac{(2\ell+1)(2\ell'+1)}{N_m(k)}\notag\\*&\qquad\times\int_0^1d\mu\,\mathcal{L}_\ell(\mu)\mathcal{L}_{\ell'}(\mu)[P(k,\mu)+P_n]^2,\label{eq:Pellcov}\end{align}
where we have used the fact that the integrand is even in $\mu$ while adapting the equations. The uncertainty in $P_\ell(k)$ is given by $\sqrt{\operatorname{cov}{[P_\ell(k),P_\ell(k)]}}$, and since the Legendre polynomials are an orthogonal basis such that
\begin{equation}
    \int_{-1}^1d\mu\,\mathcal{L}_\ell(\mu)\mathcal{L}_{\ell'}(\mu)=\frac{2}{2\ell+1}\delta_{\ell{\ell'}},
\end{equation}
taking $P_n\gg P(k,\mu)$ leads to
\begin{equation}\sigma_{P_\ell}(k)\approx\frac{P_n\sqrt{2\ell+1}}{\sqrt{N_m(k)}},\label{eq:42}\end{equation}
so that the uncertainty in the quadrupole power spectrum is
}
\begin{equation}\sigma_{P_{\ell=2}}(k) \approx \frac{P_n\sqrt{\replaced{10}{5}}}{\sqrt{N_m(k)}},\end{equation}
in the regime where the thermal noise uncertainty dominates over sample variance.

The broad implication is that if the uncertainties on the power spectra primarily come from instrumental noise, then the signal-to-noise ratio will be non-negligibly lower for the quadrupole power spectrum compared to that for the monopole. We already expect the quadrupole signal to be weaker than the monopole $P(k)$---$(4/3)\beta+(4/7)\beta^2\approx0.6$ for our illustrative value of $\beta=0.4$, and the beam-convolved shot-noise component of the quadrupole is potentially similar to the monopole shot noise over a narrow range of scales. With the noise over threefold higher, the quadrupole clustering signal will thus be an order of magnitude more challenging to detect.

\section{Models and Detectability}
\label{sec:model}

Having established the formalism around observational anisotropies in the apparent line-intensity power spectrum, and considered sources of uncertainty in measuring the power spectrum, we are ready to consider a specific scenario where the measurement of observational anisotropies in the apparent signal adds constraining power compared to measurement of the monopole alone.

Unless otherwise explicitly stated, each time $P_\ell(k)$ or $P(k,\mu)$ is used in this section (including in expressions taken from~\autoref{sec:uncertainties}), it refers to the observed signal incorporating RSD and beam smearing as opposed to the real-space signal.

\subsection{Fiducial Model and Experiment}
\label{sec:fidparams}
We consider the COMAP Phase I experiment as described in~\cite{Ihle19} and above. To calculate a prediction for $P_\text{obs}(k,\mu)$ and for $\sigma_P(k)$, we must have a fiducial $L(M)$ relation to assign a CO luminosity for a given halo mass, and any instrumental parameters that inform the noise power spectrum as expressed in~\autoref{eq:Pn}. For the former, we take a double power-law model that is similar in form to that of~\cite{Padmanabhan2018a}. However, our model has no redshift evolution and a different parametrisation, with parameter values that lead to a $L(M)$ relation that broadly matches the fiducial $L(M)$ in~\cite{Li16} (based on the halo mass--star-formation rate (SFR) relation of~\cite{Behroozi13a,Behroozi13b} and an empirical power-law scaling between SFR and CO luminosity):
\begin{equation}
    \frac{L_\text{CO}(M)}{L_\odot} = \frac{C}{(M/M_1)^A+(M/M_1)^B},
\end{equation}
with $A=-1.7$, $B=0.1$, $\log{C}=5.8$, and $\log{(M_1/M_\odot)}=12$; the minimum halo mass considered for CO emission is $10^{10}\,M_\odot$. An intrinsic log-normal scatter of $\sigma=0.4$ (in units of dex) is added to the relation. The second moment of the $L(M)$ relation then takes on an extra factor of $\exp{(\sigma^2\replaced{\log}{\ln}^2{10})}$ relative to $\sigma=0$, and since $P_\text{shot}$ is proportional to that second moment,
\begin{equation}P_\text{shot}(\sigma) = \exp{(\sigma^2\replaced{\log}{\ln}^2{10})}P_\text{shot}(\sigma=0).\end{equation}

For $P_n$, we largely use the same parameters as~\cite{Ihle19}, with some minor alterations. As in that work, we take $T_\text{sys}=40$ K and $N_\text{feeds}=19$, but we assume that instead of surveying one 2.25 deg$^2$ patch for 6000 hours, we survey two patches of $\Omega_\text{surv}=2.56$ deg$^2$ for $t_\text{surv}=3000$ hours each. This will essentially double $P_n$ in comparison to~\cite{Ihle19}, but the doubled survey volume will mean that we actually sample double the number of modes calculated in~\autoref{eq:Nmkmu} and~\autoref{eq:Nmk}. Therefore, for the purposes of this section, with $N_m$ calculated per patch,
\begin{align}\sigma_{P_{\ell=0}}(k) &= \frac{P_{\ell=0}(k)+P_n}{\sqrt{2N_m(k)}};\\
\sigma_P(k,\mu) &= \frac{P(k,\mu)+P_n}{\sqrt{2N_m(k,\mu)}};\\
\sigma_{P_{\ell=2}}(k) &\approx \frac{P_n\sqrt{\replaced{10}{5}}}{\sqrt{2N_m(k)}}.\end{align}

We use the \texttt{lim}\footnote{\url{https://github.com/pcbreysse/lim}} package to generate analytic power spectra given our fiducial CO model and the halo mass function fit of~\cite{Tinker08}. At $z=2.8$, this results in an average CO line temperature of $\avg{T}\approx0.75$ $\mu$K, a luminosity-averaged CO bias of $b=2.7$ (meaning $\beta\approx0.36$, close to our illustrative value of 0.4), and a shot-noise power spectrum of $P_\text{shot}\approx290$ $\mu$K$^2$ Mpc$^3$.

\subsection{Power Spectra and Sensitivities}
\begin{figure}[t!]
 \centering\includegraphics[width=0.84\columnwidth]{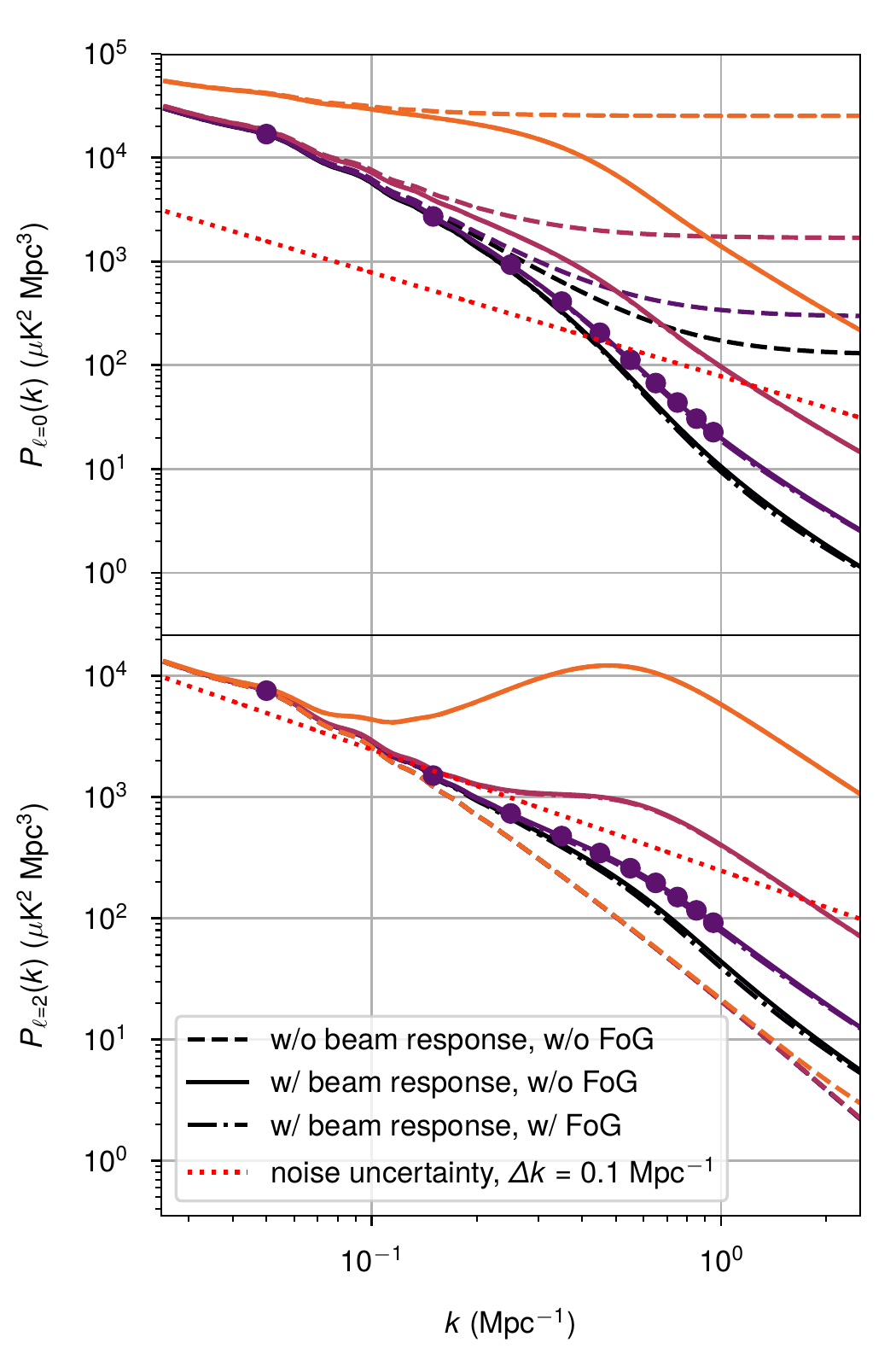}
 \caption{Monopole and quadrupole power spectra for the fiducial CO model parameters. We also vary the value of $\sigma$ from the fiducial value of 0.4 dex in order to illustrate how the power spectra change with less or more scatter. The filled circles indicate the mock data points used for Fisher matrix analysis. The red dotted line in each panel indicates uncertainties on the power spectrum from noise only, for $k$-bins of width $\Delta k=0.1$ Mpc$^{-1}$.}
 \label{fig:Pk}
\end{figure}

We calculate the following power spectra, and plot them in~\autoref{fig:Pk}: $P_{\ell=0}(k)$ and $P_{\ell=2}(k)$ from~\autoref{eq:Kaiser1} and~\autoref{eq:Kaiser2}, before instrumental anisotropies; $P_{\text{conv},\ell=0}(k)$ and $P_{\text{conv},\ell=2}(k)$ from~\autoref{eq:Kaiserconv1} and~\autoref{eq:Kaiserconv2}, accounting for beam response with the coefficients of Equations~\ref{eq:Kaiserconv3_} through~\ref{eq:Kaiserconv6_}; and the monopole and quadrupole derived from the $P_\text{obs}(k,\mu)$ of~\autoref{eq:Pk_damped}, accounting for both beam response and the small-scale damping from the `finger-of-God' (FoG) effect. We also vary the shot noise by varying $\sigma$ from the fiducial value, in order to demonstrate the effect of decreased or increased shot noise on the monopole and quadrupole signals. We also plot in~\autoref{fig:Pk} the uncertainty from instrumental noise alone when measuring these power spectra in $k$-bins of width $\Delta k=0.1$ Mpc$^{-1}$.

The plot shows that a measurement of the quadrupole should corroborate the monopole measurement of the clustering of line emission at the lowest $k$, and at the appropriate scale ($k\sim0.5$ Mpc$^{-1}$ in the case of COMAP) becomes a direct measurement of the shot noise in some cases. Note that the FoG correction is nowhere near as significant as the correction due to beam response. The damping only applies to the clustering component of the power spectrum and only becomes significant near $k\sim1$ Mpc$^{-1}$, at which point the shot-noise component already dominates the total signal. Therefore, for the remainder of this work, we will neglect the FoG correction, which allows us to use the closed-form expressions of Equations~\ref{eq:Kaiserconv3_} through~\ref{eq:Kaiserconv6_} to evaluate the monopole and quadrupole power spectrum.

We can, as in previous works like~\cite{Li16}, quote a single signal-to-noise ratio across all scales as
\begin{equation}
    (\mathrm{S/N})_{P_\ell(k)} = \left[\sum_k\frac{P_\ell^2(k)}{\sigma^2_{P_\ell}(k)}\right]^{1/2}.
\end{equation}
Summing over all $k<1$ Mpc$^{-1}$, COMAP Phase I could expect to detect the monopole $P_{\ell=0}(k)$ with $\mathrm{S/N}\approx9.3$ (or $\approx12$ omitting sample variance), and the quadrupole $P_{\ell=2}(k)$ with $\mathrm{S/N}\approx\replaced{2.3}{3.4}$.

If we define a signal-to-noise ratio for $P(k,\mu)$ analogously as
\begin{equation}
    (\mathrm{S/N})_{P(k,\mu)} = \left[\sum_k\sum_\mu\frac{P^2(k,\mu)}{\sigma^2_{P}(k,\mu)}\right]^{1/2},
\end{equation}
then this ratio over all $|\mu|<1$ and all $k<1$ Mpc$^{-1}$ is $\approx9.7$. This suggests that the monopole and quadrupole together account for the bulk of the total signal-to-noise of the anisotropic power spectrum, and thus the bulk of the information content.

\subsection{Fisher Matrix Analysis}

To quantify the information content of our observables, we undertake an analysis in the Fisher matrix formalism\footnote{See Section 11.4 of~\cite{Dodelson} for a pedagogical overview of Fisher analyses.}. While we laid out a double power-law model in~\autoref{sec:fidparams} to obtain fiducial expectations for the signal, power spectrum measurements constrain the $L(M)$ model parameters somewhat opaquely with all sorts of parameter degeneracies. We therefore focus on the parameters behind the real-space signal of~\autoref{eq:Pk_realspace}: the average line temperature $\avg{T}$, the luminosity-averaged bias $b$, and the shot-noise component $P_\text{shot}$ of the power spectrum. We fix the underlying matter power spectrum $P_m(k)$, as well as all cosmological and nuisance parameters (including $\sigma_\perp$). All this should still allow a demonstration---even if only a purely illustrative one---of the additional constraining power that the quadrupole power spectrum can provide.

We calculate the Fisher matrix across the power spectrum parameters $\{\lambda_i\}=\{\avg{T}/{\mu\mathrm{K}},b,P_\text{shot}/(100\,{\mu\mathrm{K}}^2\text{ Mpc}^3)\}$. \replaced{T}{In the simplest case where the observable quantities involved are statistically independent of each other, t}he matrix elements $F_{ij}$ are given by
\begin{equation}
    F_{ij} = \sum_k \frac{1}{\sigma^2[O_k]}\frac{dO_k}{d\lambda_i}\frac{dO_k}{d\lambda_j},
\end{equation}
where the observable vector $O_k$ is one of three possibilities:
\begin{itemize}
    \item the monopole power spectrum in $k$-bins of $\Delta k=0.1$ Mpc$^{-1}$ up to $k=1$ Mpc$^{-1}$ (the filled circles in the upper panel of~\autoref{fig:Pk});
    \item both the monopole and the quadrupole power spectrum, again in $k$-bins of $\Delta k=0.1$ Mpc$^{-1}$ up to 1 Mpc$^{-1}$ (the filled circles across both panels of~\autoref{fig:Pk});
    \item and the full $P(k,\mu)$ in the same $k$-bins and in $\mu$-sectors of $\Delta\mu=0.01$.
\end{itemize}
In \replaced{all}{the first and third} cases, we expect noise-derived uncertainties to be high enough in this particular scenario that any off-diagonal covariances between the observables may be safely ignored.\added{ However, for the second case, we factor in the covariance between the monopole and the quadrupole, which turns out to be only one order of magnitude below the noise-derived uncertainties in this scenario. We adapt the Fisher matrix formalism in this case from~\cite{Taruya2011}. Take $C_{\ell{\ell'}}(k)\equiv\operatorname{cov}{[P_\ell(k),P_{\ell'}(k)]}$ from~\autoref{eq:Pellcov} as a $2\times2$ covariance matrix between $P_{\ell=0}$ and $P_{\ell=2}$ at fixed $k$, and take $C_{\ell{\ell'}}^{-1}(k)$ to denote the inverse of that matrix. Then the monopole-plus-quadrupole Fisher matrix over $\{\lambda_i\}$ is given by
\begin{equation}
    F_{ij} = \sum_k \sum_{\ell,\ell'}\frac{dP_{\ell}(k)}{d\lambda_i}C_{\ell{\ell'}}^{-1}(k)\frac{dP_{\ell'}(k)}{d\lambda_j},
\end{equation}
where, for our purposes, the sum over $\ell$ and $\ell'$ only spans the monopole and quadrupole ($\ell=0$ and $\ell=2$).}

\begin{figure}[t!]
 \centering\includegraphics[width=\columnwidth]{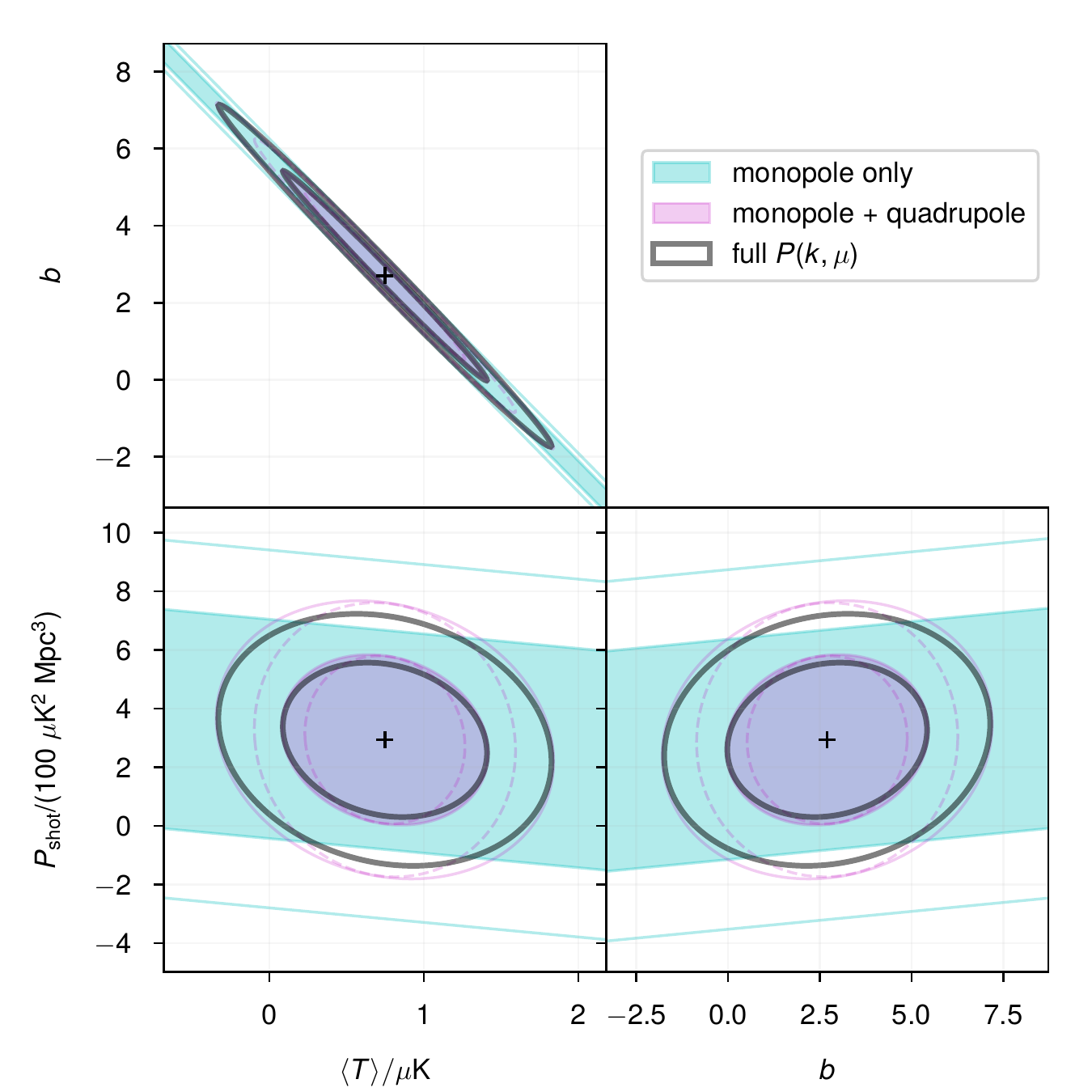}
 \caption{$1\sigma$ and $2\sigma$ error ellipses for the power spectrum parameters, given the monopole power spectrum only (cyan), both the monopole and quadrupole power spectra (magenta), and the full two-dimensional $P(k,\mu)$ (grey), all up to $k=1$ Mpc$^{-1}$. Dashed ellipses show constraints from the monopole and quadrupole if they are (wrongly) considered statistically independent.}
 \label{fig:Fisher}
\end{figure}

We show the resulting error ellipses (see~\citealt{Coe2009} for the calculations used) in~\autoref{fig:Fisher}. With the monopole measurement alone, significant parameter degeneracies exist in the constraints. The degeneracy between $\avg{T}$ and $b$ is unsurprising given that they each scale the clustering component of the real-space signal in exactly the same way; while increasing $b$ does decrease the enhancement of the signal from the Kaiser effect, the net effect of increasing $b$ will still be to increase the redshift-space signal (at least for $b>-1/3$). Incorporating the quadrupole into this analysis provides significant constraining power that is not completely codirectional with the monopole constraints, breaking this degeneracy.\added{ We also show what errors would be claimed if we incorrectly considered the monopole and quadrupole to be statistically independent, i.e.,~neglected their covariance at each $k$. The results are clearly unreasonable by virtue of outperforming even constraints from the full $P(k,\mu)$, which has information beyond the monopole and quadrupole.}

\replaced{I}{On the other hand, i}t also appears that using the full information of $P(k,\mu)$ does not improve constraints significantly beyond the combination of the monopole and quadrupole, confirming that the bulk of the information content is in those two lowest-order moments of $P(k,\mu)$.

Given that this is for a $\sim9\sigma$ detection of the monopole accompanied by a $\sim\replaced{2}{3}\sigma$ detection of the quadrupole, the relative constraining powers of the two observables will change depending on the relative strengths of the clustering and shot-noise components, and on the signal-to-noise ratio for the observables. For instance, a strong detection of the monopole power spectrum alone may characterise its shape well enough to constrain the shot-noise and clustering components separately, even at large scales. However, this specific scenario appears to show that there are regimes in which the quadrupole power spectrum provides additional constraining power beyond the monopole.

For completeness,~\autoref{sec:Fisher_classic} considers results of a Fisher analysis using the five parameters of the $L(M)$ model, including the scatter $\sigma$.

\section{Discussion}
\label{sec:discussion}

\subsection{Foregrounds and Systematics}
\label{sec:moresys}
The present work has assumed an achromatic, axisymmetric beam and no foregrounds beyond the thermal noise of the instrument. However, observers must deal with a wide array of both astrophysical foregrounds and instrumental systematics. The former category includes not only projected foreground interloper emission as~\cite{Cheng16} and~\cite{LidzTaylor16} considered, but also residuals from removal of Galactic continuum emission and point sources, which will be imbued with instrumental anisotropies just the same.

Beam variations are not the only example of the latter category---spectrometers are subject to gain variations and undesirable reflections. Consider a systematic, such as a standing wave, that varies only in the spectral direction. The anisotropic power spectrum of this systematic by itself will be entirely confined to $\mu=1$, and otherwise a function of $k_\parallel=k\mu$, which is simply equal to $k$ when $\mu=1$:
\begin{equation}
    P_\text{sys}(k,\mu) = P_\text{sys,1D}(k)\delta(\mu-1).
\end{equation}
We may then calculate the monopole and quadrupole of this systematic by itself:
\begin{align}
    P_{\text{sys},\ell=0}(k,\mu)&= P_\text{sys,1D}(k);\\
    P_{\text{sys},\ell=2}(k,\mu)&= 5P_\text{sys,1D}(k).
\end{align}
Therefore, standing waves and other purely spectral systematics could be five times more powerful in the quadrupole than in the monopole, which is not true of the CO signal. In the case of COMAP, standing waves may be produced in a range of optical cavities, from the 0.5 m cavity between the COMAP amplifiers and receiver window to the 2 m cavity between the receiver and the secondary mirror. A standing wave in an air cavity of physical length $L$ will have a spectral periodicity of $\Delta\nu_\text{SW}=c/(2L)$, which then corresponds to a comoving wavenumber of
\begin{align}
    k_\text{SW} = 2\pi\left(\frac{c(1+z)^2}{H(z)}\frac{\Delta\nu_\text{SW}}{\nu_\text{rest}}\right)^{-1} = \frac{4\pi H(z)}{c(1+z)^2}\frac{L\nu_\text{rest}}{c}.
\end{align}
For $L=0.5$ to 2 m, $k_\text{SW}$ ranges from 0.16 to 0.63 Mpc$^{-1}$, which lies squarely in the range where we expect significant enhancement of the quadrupole given~\autoref{fig:Pconv_coeffs}.

No part of this discussion should discourage single-dish experiments from attempting a measurement of the quadrupole, however. In this simplest example, the standing wave clearly behaves very differently in the monopole and quadrupole compared to the signal of interest. Attempting to subtract the standing wave might still result in unacceptably high residuals, but excising a relatively narrow span of $(k,\mu)$ would be sufficient to remove purely spectral systematics with minimal loss of sensitivity.

As for more complex systematics or astrophysical foregrounds subjected to instrumental anisotropies, the extensive literature on Fourier-space foreground mitigation strategies developed for 21 cm intensity mapping~\citep{MW2010,Morales2012,Liu2014a,Liu2014b,Switzer15} should be applicable to single-dish surveys of other spectral lines. While many of these works deal with interferometric 21 cm intensity mapping, the work of~\cite{Switzer15} examines continuum foregrounds seen through single-dish instrumental response, in the context of 21 cm at $z\sim1$ as observed previously in pioneering work using the Robert C.~Byrd Green Bank Telescope~\citep{Chang10,Masui13,Switzer13}. Their work is thus particularly applicable to CO and [\ion{C}{2}] surveys, which are predominantly single-dish observations.

\subsection{Implications for Cross Shot Noise}
We have concerned ourselves entirely with the auto power spectrum of a single spectral line. However, the line-intensity mapping community has significant interests in using cross-correlations against galaxy surveys to find, for example, the neutral hydrogen content of different types of galaxies~\citep{Wolz17} or the CO luminosities of active galactic nuclei~\citep{BreysseAlexandroff19}. In these particular cases, the shot-noise component of the line-galaxy cross power spectrum is a direct measure of the mean line brightness of the galaxy sample. Cross shot noise between intensities of two lines, with appropriate interpretation, probes the interstellar medium in high-redshift galaxies, and can even be done within the same survey (for example, between $^{12}$CO and $^{13}$CO in COMAP---see~\citealt{BreysseRahman17}).

The effect of instrumental anisotropies on the full cross power spectrum of two tracers are simply largely beyond the scope of the present work. We may readily see that, in the case of two tracers with similar bias $b$ covering the same redshift range,~\autoref{eq:Kaiser1} and~\autoref{eq:Kaiser2} apply as is (and with matched filtering, even the rest of~\autoref{sec:formalism} may apply without modification). However, we leave the calculation of the clustering cross power spectrum between tracers of wildly different biases to future work by others.

That said, we can readily show that the same algebraic results for the auto shot noise apply to the cross shot noise, with appropriate re-definitions of $\sigma_\parallel$ and $\sigma_\perp$. Suppose we observed tracer 1 (for instance, the line brightness temperature) with a survey that imposes a given $\sigma_{\perp,1}$ and $\sigma_{\parallel,1}$, and the survey of tracer 2 (for instance, the galaxy density contrast) has its own characteristic associated $\sigma_{\perp,2}$ and $\sigma_{\parallel,2}$. Then the Fourier transform of the individual tracers $T_1(\mathbf{x})$ and $T_2(\mathbf{x})$ in convolution with their respective instrument responses are given as in~\autoref{eq:Tconv}:
\begin{align}\tilde{T}_{1,\text{conv}}(\mathbf{k})&=\tilde{T}_1(\mathbf{k})\exp{(-k_\perp^2\sigma_{\perp,1}^2/2-k_\parallel^2\sigma_{\parallel,1}^2/2)};\\\tilde{T}_{2,\text{conv}}(\mathbf{k})&=\tilde{T}_2(\mathbf{k})\exp{(-k_\perp^2\sigma_{\perp,2}^2/2-k_\parallel^2\sigma_{\parallel,2}^2/2)}.\end{align}
Then the `true' cross power spectrum is given by
\begin{equation}
    P_{1\times2}(\mathbf{k})=\operatorname{Re}{(\tilde{T}_1^*\tilde{T}_2)},
\end{equation}
and the cross power spectrum after instrumental response by
\begin{align}
    P_{1\times2,\text{conv}}(\mathbf{k})&=\operatorname{Re}{(\tilde{T}_{1,\text{conv}}^*\tilde{T}_{2,\text{conv}})}
    \\&=P_{1\times2}(\mathbf{k})\notag\\*&\quad\times\exp{\left[-\frac{k_\perp^2(\sigma_{\perp,1}^2+\sigma_{\perp,2}^2)}{2}-\frac{k_\parallel^2(\sigma_{\parallel,1}^2+\sigma_{\parallel,2}^2)}{2}\right]}.
\end{align}
Then we can define
\begin{align}
    \sigma_{\perp,1\times2}&=\frac{(\sigma_{\perp,1}^2+\sigma_{\perp,2}^2)^{1/2}}{\sqrt{2}},\\\sigma_{\parallel,1\times2}&=\frac{(\sigma_{\parallel,1}^2+\sigma_{\parallel,2}^2)^{1/2}}{\sqrt{2}};
\end{align}
these can then be used as $\sigma_\parallel$ and $\sigma_\perp$ for~\autoref{eq:Kaiserconv4},~\autoref{eq:Kaiserconv6},~\autoref{eq:Kaiserconv4_}, and~\autoref{eq:Kaiserconv6_}. For example, if we cross-correlate COMAP with a galaxy survey with very fine angular and redshift resolution, this effectively decreases $\sigma_\parallel$ by a factor of $\sqrt{2}$ in cross-correlation compared to in auto-correlation. The wavenumbers at which $S_0$ and $S_2$ rise and fall merely shift up by a factor of $\sqrt{2}$. This does mean that we expect the cross shot noise to add significantly to the cross quadrupole power spectrum, just over a slightly different range of $k$ compared to the auto quadrupole. Since the cross-correlation exercise will also reject disjoint systematics and reduce uncertainties from noise, errors on the cross monopole may be small enough to make measuring the cross quadrupole a somewhat redundant exercise. Whether the cross quadrupole adds significant information surrounding the cross shot noise should be evaluated on a case-by-case basis in future work that we leave to others.
\section{Conclusions}
\label{sec:conclusions}

We have considered multiple important observational anisotropies that distort the line-intensity signal targeted by a survey with limited resolution. In particular, we explicitly consider the effect of instrumental resolution limits on the signal, rather than folding it into the uncertainties on the signal. By calculating the resulting quadrupole power spectra in addition to $P(k,\mu)$, we clarify the form of the distorted signal, and in particular the fact that shot noise will contribute to the quadrupole due solely to instrumental anisotropies, and at scales relevant to single-dish line-intensity surveys.

We emphasise that the results of this work should not be seen as a call for line-intensity mappers with infinitely poor spectral or spatial resolution. In the end, resolution limits do attenuate the signal---it would be far more preferable to access redshift-space signals with infinitely fine resolution (as represented by the dashed curves of~\autoref{fig:Pk}). However, through explicit consideration of the distorted signals, we have shown that attenuation of the monopole does not preclude recovery of some information via measurement of the quadrupole. As a full understanding of instrumental anisotropies is crucial in such recovery, instruments and surveys should plan to oversample in frequency and on the sky to characterise the instrumental response in all dimensions \emph{in situ}, as applicable. The hope is that future work in the broader line-intensity mapping community, following the lead of the 21 cm intensity mapping community, looks beyond forecasts with the ideal real-space $P(k)$ and takes careful inventory of observational anisotropies, as these suggest approaches to measuring the line-intensity signal that the real-space $P(k)$ alone does not.

\acknowledgements{Thanks to Sarah Church (my doctoral advisor), Risa Wechsler (and her group), Eiichiro Komatsu, Patrick Breysse, Hamsa Padmanabhan, and other members of the COMAP and CCAT-prime collaborations for discussions informing this work at its various stages. Special thanks go to P.~Breysse and H.~Padmanabhan for their careful readings of this work close to submission\added{, and to the anonymous referee whose comments significantly improved this manuscript}. This research made use of NASA's Astrophysics Data System Bibliographic Services.}

\software{\texttt{hmf}~\citep{hmf}; Matplotlib~\citep{matplotlib}; Astropy, a community-developed core Python package for astronomy~\citep{astropy}.}

\appendix
\section{Calculation of Hexadecapole-related Quantities}
\label{sec:hexadecapole}
The Kaiser formula as used in~\autoref{eq:Pk_Kaiser} not only predicts a non-zero quadrupole ($\ell=2$), but also results in a non-vanishing hexadecapole, which we may find by setting $\ell=4$ and substituting~\autoref{eq:Pk_Kaiser} into~\autoref{eq:legendre}:
\begin{equation}
    P_{\ell=4}(k) = \frac{9}{8}\int_0^1d\mu\,P(k,\mu)(35\mu^4-30\mu^2+3) = \frac{8}{35}\beta^2\avg{T}^2b^2P_m(k).
\end{equation}

For our illustrative value of $\beta=0.4$, the redshift-space hexadecapole is less than 4\% of the real-space power spectrum, and less than 3\% of the redshift-space $P_0(k)$---an order of magnitude weaker than the quadrupole. This is admittedly before instrument response introduces its own anisotropies, so if we define $C_4$ and $S_4$ such that
\begin{equation}
    P_{\text{conv,}\ell=4}=C_4\avg{T}^2b^2P_m(k)+S_4P_\text{shot},
\end{equation}
we obtain, again defining $\delta_\alpha^2=\sigma_\parallel^2-\sigma_\perp^2$ as in the main text,
\begin{align}
    C_4&=\frac{9}{256}\left\{\frac{3\pi^{1/2}\exp{(-\alpha_\perp^2)}\erf{[(\delta_\alpha^2)^{1/2}]}}{(\delta_\alpha^2)^{9/2}}\left[1225\beta^2-100\delta_\alpha^2(3\beta-7)\beta+4\delta_\alpha^4(35-60\beta+3\beta^2)+16(\beta-5)\delta_\alpha^6+16\delta_\alpha^8\right]\right.\notag\\*&\left.\hspace{4cm}-\frac{2\exp{(-\alpha_\parallel^2)}}{\delta_\alpha^8}\left[3675\beta^2+50\delta_\alpha^2\beta(42+31\beta)+4\delta_\alpha^4(105+170\beta+104\beta^2)+8\delta_\alpha^6(5+16\beta+8\beta^2)\right]\right\};\label{eq:Kaiserconv7}\\
    S_4&=\frac{9}{64}\left[\frac{3\pi^{1/2}\exp{(-\alpha_\perp^2)}\erf{[(\delta_\alpha^2)^{1/2}]}}{(\delta_\alpha^2)^{5/2}}(35-20\delta_\alpha^2+4\delta_\alpha^4)-\frac{10\exp{(-\alpha_\parallel^2)}}{\delta_\alpha^4}(21+2\delta_\alpha^2)\right].\label{eq:Kaiserconv8}
\end{align}
In the limit of $\sigma_\parallel\to0$, we obtain
\begin{align}
    C_4&=\frac{9}{128\alpha_\perp^9}\left\{8\alpha_\perp^7(5+16\beta+8\beta^2)-4\alpha_\perp^5(105+170\beta+104\beta^2)+50\alpha_\perp^3\beta(42+31\beta)-3675\alpha_\perp\beta^2\right.\notag\\*&\left.\hspace{2cm} +3[1225\beta^2+100\alpha_\perp^2(3\beta-7)\beta+4\alpha_\perp^4(35-60\beta+3\beta^2)-16(\beta-5)\alpha_\perp^6+16\alpha_\perp^8]F(\alpha_\perp)\right\};\\
    S_4&=\frac{9}{32\alpha_\perp^5}[5\alpha_\perp(2\alpha_\perp^2-21)-(12\alpha_\perp^2(\alpha_\perp^2+5)+105)F(\alpha_\perp)].
\end{align}

\begin{figure}[t!]
 \centering\includegraphics[width=0.64\columnwidth]{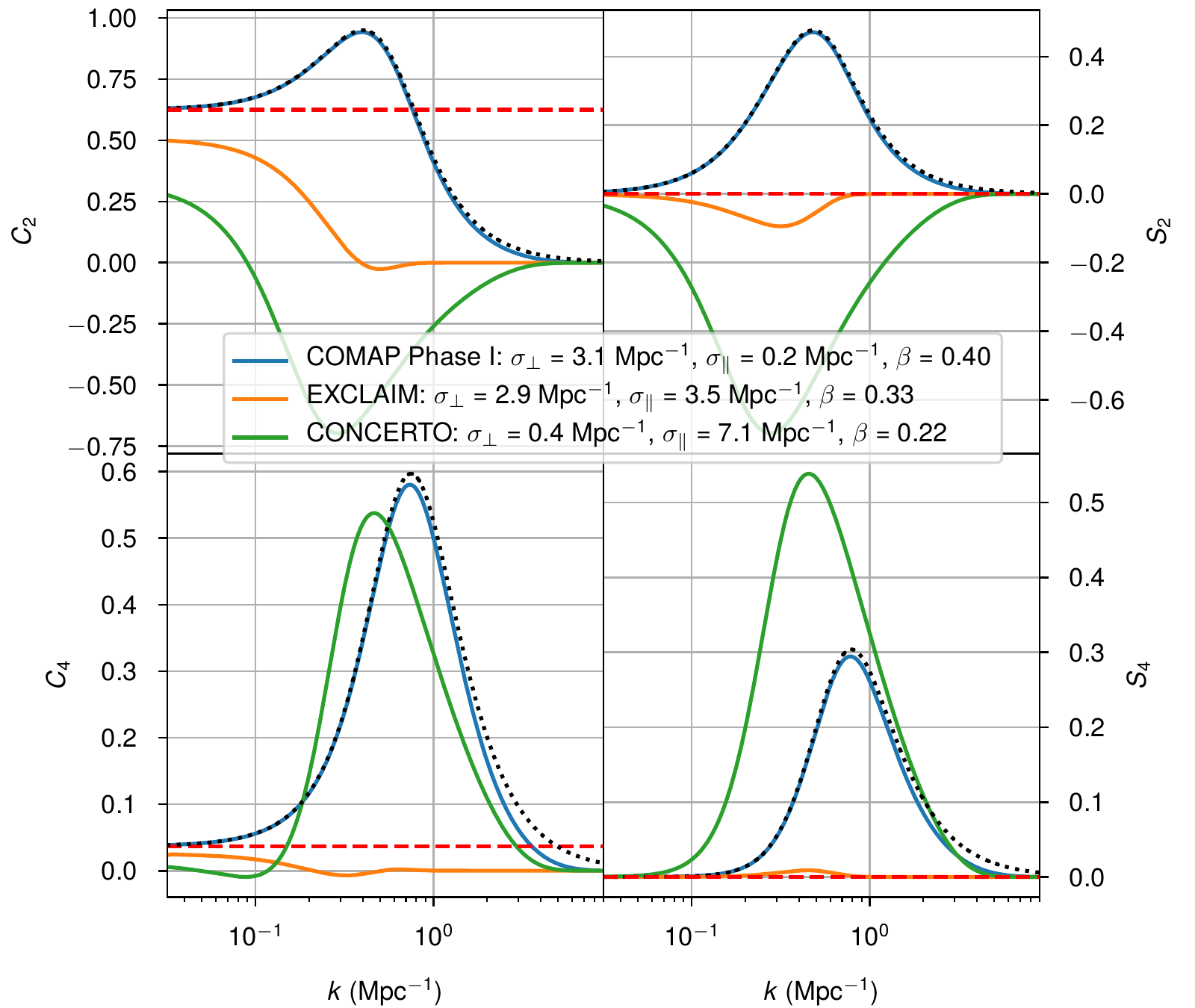}
 \caption{The coefficients $C_2$, $S_2$, $C_4$, and $S_4$ of Equations~\ref{eq:Kaiserconv5},~\ref{eq:Kaiserconv6},~\ref{eq:Kaiserconv7}, and~\ref{eq:Kaiserconv8} as functions of $k$, given the example values of $\sigma_\perp$ and $\sigma_\parallel$ for the surveys of~\autoref{tab:examplesigmas}, and illustrative values of $\beta$ loosely based on~\cite{MDK19}. The value of $\sigma_\parallel$ in COMAP Phase I has been chosen to match $1/4$ of the nominal channel width, to reflect the higher intrinsic frequency resolution of the spectrometer. Black dotted lines in each panel show the limit of the coefficients for COMAP Phase I as $\sigma_\parallel\to0$. Red dashed lines in each panel show the limit given $\beta=0.4$ as $\sigma_\perp\to0$ and $\sigma_\parallel\to0$.}
 \label{fig:hexadecapole_coeffs}
\end{figure}
We show values of $C_4$ and $S_4$ in~\autoref{fig:hexadecapole_coeffs}, re-plotting the quadrupole analogues $C_2$ and $S_2$ from~\autoref{fig:Pconv_coeffs} alongside for comparison. The instrumental anisotropies significantly boost the hexadecapole at intermediate ranges of $k$, as much as to the same order of magnitude as the quadrupole.

The uncertainties from noise, however, are greater than for the quadrupole. \replaced{As with the quadrupole in the main text, consider a data cube of Gaussian noise with randomly distributed $P(\mathbf{k})$ with mean $P_n$ and standard deviation $P_n$, with an uncorrelated value of $\mu\in(0,1)$ associated with each value of $P(\mathbf{k})$. Then since
\begin{equation}
    \sigma^2[7\mu^4-6\mu^2] = \avg{(7\mu^4-6\mu^2)^2}-\avg{7\mu^4-6\mu^2}^2 = 29/45-9/25
    = 64/225,
\end{equation}
we obtain that for this noise cube,
\begin{equation}
    \sigma^2[(35\mu^4-30\mu^2+3)P(\mathbf{k})] = \sigma^2[35\mu^4-30\mu^2+3]\left\{\sigma^2[P(\mathbf{k})]+\avg{P(\mathbf{k})}^2\right\}=\frac{128}{9}P_n^2.
\end{equation}
(Note that as in~\autoref{eq:42}, we have implicitly simplified the expression based on the fact that $\avg{35\mu^4-30\mu^2+3}=0$.) Then, i}{I}n the regime where uncertainty from such Gaussian instrument noise dominates over sample variance,\added{ applying~\autoref{eq:42} gives}
\begin{equation}
    \sigma_{P_{\ell=4}}(k)\approx\frac{\replaced{3\sqrt{2}}{3}P_n}{\sqrt{N_m(k)}}.
\end{equation}
For the quadrupole, we expect a signal roughly half that of the monopole (before beam response mixes shot noise into the quadrupole) with noise uncertainties higher by $\sqrt{\replaced{10}{5}}$, meaning that the ratio of the signal to the instrumental noise uncertainty alone (excluding sample variance) should be approximately \replaced{six}{four} times lower for the quadrupole than for the monopole. For the hexadecapole, we expect a signal somewhere between 3\% \added{of the monopole} and the same order of magnitude as the quadrupole, and almost $\sqrt{2}$ times the\added{ quadrupole} noise uncertainty, meaning the hexadecapole signal-to-noise ratio is unlikely to exceed one-half of the \emph{quadrupole} signal-to-noise ratio (or one-twelfth of the monopole signal-to-noise ratio).

For the example of COMAP Phase I considered in~\autoref{sec:model} of the main text, the expected signal-to-noise ratio for the quadrupole is already marginal at \replaced{2.3}{around 3}, which suggests that even a tentative detection of the hexadecapole is somewhat unlikely. This would explain the relatively small gain in constraining power when using the full $P(k,\mu)$ (which should include any information gain from the next-to-leading anisotropies represented by the hexadecapole) relative to using just the monopole and quadrupole power spectra.

\section{Fisher Matrix Analysis with Double Power-law Parameters}
\label{sec:Fisher_classic}
While we use the parameters of~\autoref{eq:Pk_realspace} (taking $P_m(k)$ to be fixed) in the Fisher analysis of the main text, it is possible to take the five parameters of our double power-law parametrisation, $\{\lambda_i\}=\{A,B,\log{C},\log{(M_1/M_\odot)},\sigma\}$, and calculate the Fisher matrix across these five parameters. As it turns out, however, the Fisher matrix is ill-conditioned without some informative priors. We therefore include finite prior widths for the power-law slopes and $\log{(M_1/M_\odot)}$ around the fiducial parameter values---namely, $\sigma_\text{prior}[A]=2$, $\sigma_\text{prior}[B]=1$, and $\sigma_\text{prior}[\log{(M_1/M_\odot)}]=2$. We have a reasonable expectation for the $L_\text{CO}(M)$ slopes from abundance matching and empirical modelling (as in \citealt{Padmanabhan2018a}; see also the compilation of models in~\citealt{Li16}), and~\cite{Behroozi13a} expect a decline in star-formation efficiency beyond $M_\text{vir}\sim10^{12}\,M_\odot$, for which they find support in previous work.

\begin{figure}[t!]
 \centering\includegraphics[width=0.64\columnwidth]{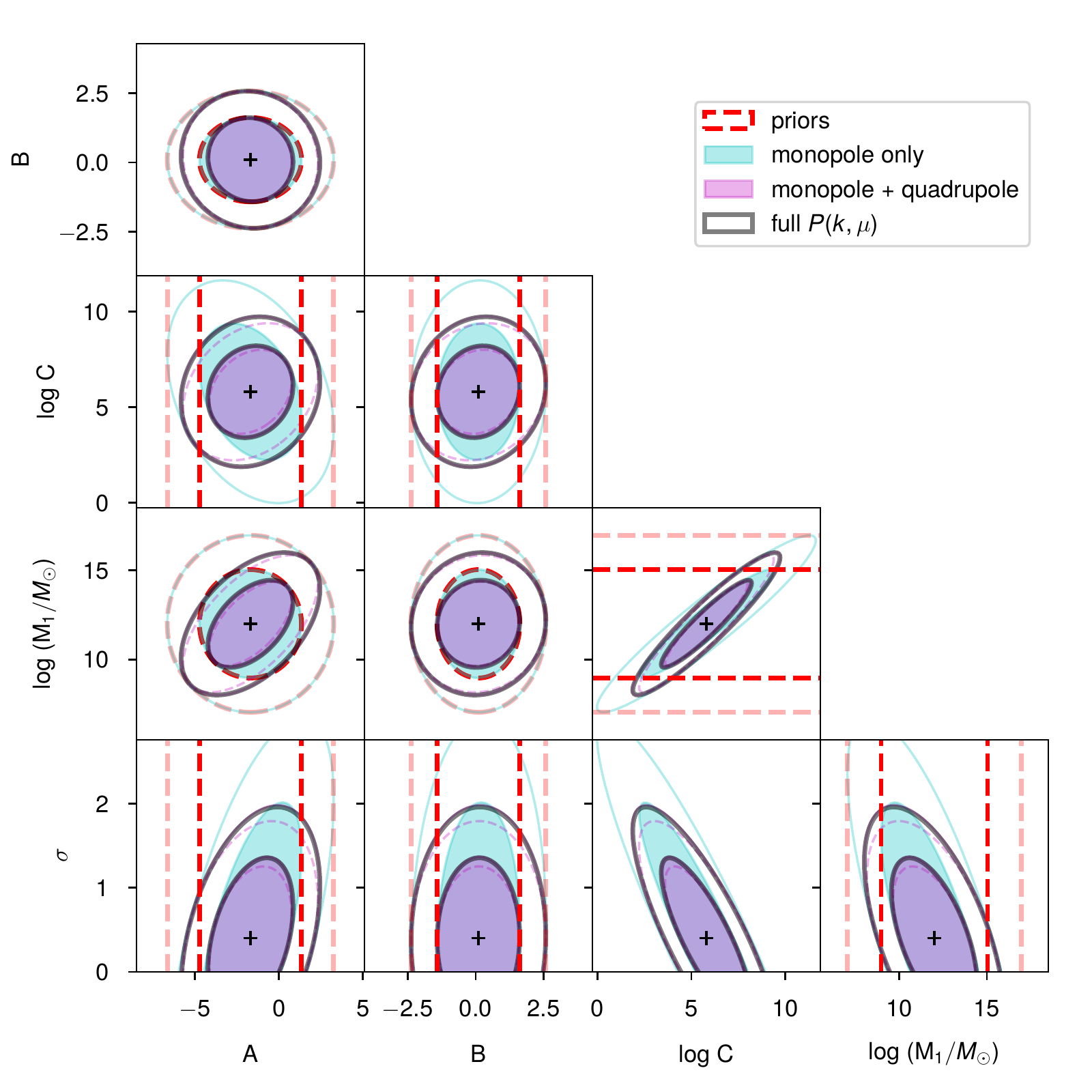}
 \caption{$1\sigma$ and $2\sigma$ error ellipses for the CO model parameters, given the monopole power spectrum only (cyan), both the monopole and quadrupole power spectra (magenta), and the full two-dimensional $P(k,\mu)$ (grey), all up to $k=1$ Mpc$^{-1}$. Priors (red dashed) are also shown. Magenta dashed ellipses show constraints from the monopole and quadrupole if they are (wrongly) considered statistically independent.}
 \label{fig:Fisher_classic}
\end{figure}

We show the resulting error ellipses in~\autoref{fig:Fisher_classic}. The quadrupole provides significant constraining power from indirectly constraining the shot noise that is subdominant in the monopole for the scales considered, whereas in many cases the priors bound the monopole-only error ellipses. The strongest improvements are in constraining the overall normalisation $C$ and the scatter $\sigma$, which mirrors the fact that measuring the quadrupole constrains $\avg{T}$ (or $b$) against $P_\text{shot}$ in a way that measuring the monopole alone cannot. As in~\autoref{fig:Fisher}, using the full information of $P(k,\mu)$ does not improve constraints significantly beyond the combination of the monopole and quadrupole.

Note that this Fisher analysis is \emph{still} a purely illustrative exercise---even more so than the exercise in the main text, as the prior widths used will affect the results significantly. If an observer is confident enough in certain highly informative priors, the quadrupole power spectrum may not provide constraining power that exceeds those priors.\deleted{\footnote{We expect a more detailed exploration of the double power-law model in this work and priors approriate for COMAP analysis in forthcoming work by the COMAP collaboration, currently in preparation.}} Furthermore, the voxel intensity distribution (VID) of the line-intensity map would provide additional orthogonal constraints on these parameters; a mock analysis combining the monopole and quadrupole power spectra with the VID would be a natural extension of the work of~\cite{Ihle19}.\added{ We expect a more detailed exploration of the double power-law model in this work and priors appropriate for COMAP analysis in forthcoming work in preparation by the COMAP collaboration, as well as future work featuring more detailed simulations and sensitivity forecasts for redshift-space monopole and quadrupole spectra in a COMAP-specific context.}

%% The reference list follows the main body and any appendices.
%% Use LaTeX's thebibliography environment to mark up your reference list.
%% Note \begin{thebibliography} is followed by an empty set of
%% curly braces.  If you forget this, LaTeX will generate the error
%% "Perhaps a missing \item?".
%%
%% thebibliography produces citations in the text using \bibitem-\cite
%% cross-referencing. Each reference is preceded by a
%% \bibitem command that defines in curly braces the KEY that corresponds
%% to the KEY in the \cite commands (see the first section above).
%% Make sure that you provide a unique KEY for every \bibitem or else the
%% paper will not LaTeX. The square brackets should contain
%% the citation text that LaTeX will insert in
%% place of the \cite commands.

%% We have used macros to produce journal name abbreviations.
%% \aastex provides a number of these for the more frequently-cited journals.
%% See the Author Guide for a list of them.

%% Note that the style of the \bibitem labels (in []) is slightly
%% different from previous examples.  The natbib system solves a host
%% of citation expression problems, but it is necessary to clearly
%% delimit the year from the author name used in the citation.
%% See the natbib documentation for more details and options.

\bibliographystyle{aasjournal}
\bibliography{references}

%% This command is needed to show the entire author+affilation list when
%% the collaboration and author truncation commands are used.  It has to
%% go at the end of the manuscript.
%\allauthors

%% Include this line if you are using the \added, \replaced, \deleted
%% commands to see a summary list of all changes at the end of the article.
%\listofchanges

\end{document}